\documentclass[5p]{elsarticle}
\usepackage[english]{babel}
\usepackage{graphicx}
\usepackage{amsmath, amssymb, upgreek}

\begin{document}

\begin{frontmatter}

\title{Dynamics and Energy Dissipation of a Rigid Dipole Driven by the RF-field in a Viscous Fluid: Deterministic Approach}

\author{T.~V.~Lyutyy\corref{cor_auth}}
\cortext[cor_auth]{Corresponding author}
\ead{lyutyy@oeph.sumdu.edu.ua}
\address{Sumy State University, 2 Rimsky-Korsakov Street, UA-40007 Sumy, Ukraine}

\begin{abstract}
The deterministic rotation of a ferromagnetic nanoparticle in a fluid is considered. The heating arising from viscous friction of a nanoparticle driven by circularly and linearly polarized alternating magnetic fields is investigated. Since the power loss of such fields depends on the character of the induced motion of a nanoparticle, all types of particle trajectories are described in detail. The dependencies of the power loss on the alternating field parameters are determined. The optimal conditions for obtaining the maximum heating efficiency are discussed. The effect of heating enhancement by a static field is analyzed. The results obtained are actual for the description of heating in the magnetic fluid hyperthermia cancer treatment, when the size of the particles used is a few tens of nanometers.
\end{abstract}


\end{frontmatter}

\section{INTRODUCTION}
\label{Int}

Practical applications of ferrofluids \cite{Rosensweig1985Ferrohydrodynamics,0038-5670-17-2-R02} are connected to their interaction with external magnetic fields. In general case, the response of a ferrofluid to these fields is determined by the dynamics of each ferromagnetic nanoparticle included in the ferrofluid. The individual motion of the nanoparticle defines the performance of modern therapy techniques, such as targeted drug delivery \cite{0022-3727-36-13-201,VEISEH2010284} and magnetic fluid hyperthermia \cite{0022-3727-36-13-201,JORDAN1999413,0957-4484-25-45-452001}, as well as biotechnology techniques, such as biosensors, macro-molecules, and virus separation \cite{0022-3727-36-13-201,TIAN2016420,C6AY00721J}.

There are two types of dynamics occurring in a coupled way: the internal motion of the nanoparticle magnetic moment relative to the nanoparticle crystal lattice and the motion of the whole particle with respect to the surrounding fluid. Strictly speaking, both of them are stochastic and should be described statistically. However, at present, the essential progress in the description of the forced coupled motion and the energy dissipation is achieved only in the deterministic approximation \cite{0022-3727-39-22-002,doi:10.1063/1.4737126,PhysRevB.95.134447,doi:10.21272/jnep.8(4(2)).04086,Lyutyy201887}. The accounting of the thermal noise for the coupled dynamics is in the initial stage now \cite{PhysRevB.95.104430,0031-9155-63-3-035004}.

Therefore, a simplified model, where the internal magnetic dynamics is neglected, is currently widely used \cite{PhysRevE.63.011504,PhysRevE.83.021401,0953-8984-15-23-313,SotoAquino201546,0953-8984-15-23-313,PhysRevE.92.042312,PhysRevE.97.052611}. Within this framework, due to the strong anisotropy, the magnetic moment is supposed to be fixed to the nanoparticle easy axis. This approximation is also called the "rigid dipole" or the "frozen magnetic moment" model. The problem of the rigid dipole rotation about a fixed center (known as spherical motion) was considered first in \cite{doi:10.1063/1.1656014}. For today, the most complete study of the influence of thermal activation and inter-particle dipole interaction on the energy dissipation during the forced spherical motion of a rigid dipole was reported in \cite{PhysRevE.97.052611}. However, some purely dynamical effects remain unclear. Thus, the aim of the present work consists in a deep analysis of this problem in the whole range of the external field parameters. A special attention is paid to the power loss and its control. The actuality of these findings is also stimulated by practical reasons: in the magnetic fluid hyperthermia method of cancer treatment exactly the viscous rotation is the main energy dissipation channel if nanoparticles are large enough ($> 20 \mathrm{nm}$) \cite{andr2006magnetism}.

\section{DESCRIPTION OF THE MODEL}
\label{Desc}
Lets consider a ferromagnetic particle of radius $R$, uniform mass density $\rho$ and magnetization $\mathcal{M}$ placed into a fluid of viscosity $\eta$. The spherical motion, i.e. the motion with fixed center of mass, of such particle is described by the following system of equations \cite{doi:10.1063/1.1656014}:
\begin{equation}
\begin{array}{lcl}
    \dot{\boldsymbol{\upomega}}=\dfrac{1}{\tau_0^2}\mathbf{m}\times \mathbf{h}-\dfrac{1}{\tau_r}{\boldsymbol{\upomega}}, \\ [8pt]
    \dot{\mathbf{m}}=\boldsymbol{\upomega}\times \mathbf{m}, \\
    \label{eq:Main_Eq}
\end{array}
\end{equation}
where $\boldsymbol{\upomega}$ is the nanoparticle angular velocity, $\mathbf{m} = \mathbf{M}/M$ is the unit vector of the nanoparticle magnetic moment $\mathbf{M}$, $M = |\mathbf{M}| = \mathcal{M} V$, $V = \frac{3}{4} \pi R^3$ is the nanoparticle volume, $\mathbf{h} = \mathbf{H}/H_{a}$ is the reduced external field, $H_{a}$ is the effective anisotropy field, $\tau_0 = \sqrt{I/\mu_0 M H_{a}}$, and $\tau_r = I/6 \eta V$ with $I = \frac{2}{5} \rho V R^2$ been the moment of inertia of the nanoparticle, are the characteristic times, $\mu_0$ is the vacuum permittivity,  and the cross sign denotes the vector product. Here, the friction is treated in the Stokes approximation for Reynolds numbers, smaller than 10 \cite{Frenkel:106808}. It needs to note that the condition $\boldsymbol{\upomega} \bot \mathbf{m}$ holds for times $t \gg \tau_r$, since $\boldsymbol{\upomega} \cdot \mathbf{m} \sim \exp(-t/\tau_r)$.

The energy dissipation during the described rotational motion of the particle about its center occurs due to viscous friction. When an external alternating field is applied, the dissipated energy is compensated by the energy of this filed. The energy dissipation per unit time, or the power loss $Q$, can be introduced using the variation of the magnetic energy, which is generated by the magnetic moment increment $\delta \mathbf{M}$ in the external field $\mathbf{H}(t)$. If all the energy changes are transformed into the irreversible losses, one can write $\delta Q = \mathbf{H}(t)\delta \mathbf{M}$. The resulting $Q$ value is obtained by averaging over an observation time as well as in \cite{PhysRevB.91.054425}
\begin{equation}
    Q = \lim_{\tau \to \infty} \frac{1}{\tau} \int_{0}^{\tau} \mathbf{H}(t) \frac{\partial\mathbf{M}}{\partial t}dt.
    \label{eq:def_Q}
\end{equation}
In the reduced form $\widetilde{Q} = Q/(\mu_0 M H_a \Omega_{cr})$ ($\Omega_{cr} = \tau_r/\tau_0^2$ - characteristic frequency), which is the quantity of our main interest, the power loss can be written in the form
\begin{equation}
    \widetilde{Q} = \lim_{\widetilde{\tau} \to \infty} \frac{1}{\widetilde{\tau}}
    \int_{0}^{\widetilde{\tau}} \mathbf{h}(\widetilde{t})
    \frac{\partial\mathbf{m}}{\partial \widetilde{t}} d\widetilde{t},
    \label{eq:def_Q_red}
\end{equation}
where $\widetilde{t} = t \Omega_{cr}$, $\widetilde{\tau} = \tau \Omega_{cr}$. It is reasonable to underline here that in the simplest cases of periodic forced motion of $M$, the integration in Eq.~(\ref{eq:def_Q_red}) can be conducted over the reduced field period ($\widetilde{\mathcal{T}} = \mathcal{T}\Omega_{cr}$) only,
\begin{equation}
    \widetilde{Q} = \frac{1}{\widetilde{\mathcal{T}}}
    \int_{0}^{\widetilde{\mathcal{T}}} {\mathbf{h}(\widetilde{t})\frac{\partial\mathbf{m}}{\partial\widetilde{t}}} d\widetilde{t}.
    \label{eq:def_Q_red_1}
\end{equation}

\subsection{The validity of noise free approximation}
\label{Mod_App}

The used system of equations is valid when the magnetic moment is considered as "frozen" into crystal lattice. Such concept implies that the following conditions hold: $H_a \gg H$ and $\kappa \gg 1$, $\kappa =\mu_0 \mathcal{M} H V/(k T)$ (here $k= 1.38 \cdot 10^{-23} \mathrm{J}/\mathrm{K}$ is the Boltzmann constant, $T$ is the thermodynamic temperature). In accordance with the first condition, the magnetic moment does not practically deviate from the anisotropy axis due to an external field action. At the same time, the latter condition allows to treat the trajectory of $\mathbf{m}$ as approximately regular and, together with the first condition, exclude the essential deviation of the magnetic moment from the crystal axis due to thermal activation. These conditions are true for the real nanoparticles of maghemite \cite{C3RA45457F} with the determined parameters: average radius $R=20\mathrm{nm}$, $H_a = 7.24\cdot10^{4}\mathrm{A/m}$ and $\mathcal{M} = 3.38\cdot10^{5} \mathrm{A/m}$. Thus, when $H = 0.05 H_{a}$, $\kappa \approx 12$.

Besides the requirements to the external field amplitude, the corresponding restrictions to the field frequency exist. First, frequency is bounded below according to the following. Even if $\kappa \gg 1$, the significant changes of the angular coordinates can occur due to thermal excitation, when the observation time is much more than the time of Brownian relaxation \cite{0038-5670-17-2-R02} $\tau_B = 3 \eta V/(k T)$. It imposes the existence of the character frequency  $\Omega_{B} = 1/\tau_B = k T/(3 \eta V)$. Further, the rigid dipole model can be violated during the N\'{e}el relaxation time $\tau_N = (\Gamma/\pi)^{-1/2}\exp(\Gamma)(2\alpha' \gamma \mu_0 H_{a})^{-1}$ (here $\alpha' \ll 1$ is the dimensionless damping constant, $\gamma \thickapprox 1.76 \cdot 10^{11} \mathrm{rad}\cdot \mathrm{s}^{-1}\cdot \mathrm{T}^{-1}$ is the gyromagnetic ratio) \cite{PhysRev.130.1677,DENISOV1998282}, $\Gamma=\mu_0 \mathcal{M} H_{a}V/(k T)$. This fact causes the existence of one more character frequency $\Omega_{N} = 1/\tau_N = 2\alpha' \gamma \mu_0 H_{a}(\Gamma/\pi)^{1/2}\exp(-\Gamma)$. Thus, one can conclude that $\Omega \gg \mathrm{max}[\Omega_{B}, \Omega_{N}]$. For the above mentioned maghemite nanoparticles \cite{C3RA45457F}, $\alpha' = 0.01$ and $T=310 \mathrm{K}$, $\eta = 5\cdot10^{-3} \mathrm{Pa}\cdot\mathrm{s}^{1}$ (corresponds to blood), one can write $\Omega_B \approx 8.54\cdot10^{3} \mathrm{Hz}$ and $\Omega_N \approx 2.08\cdot10^{-96} \mathrm{Hz}$ or, finally, $\Omega \gg 8.54\cdot10^{3} \mathrm{Hz}$.

At the same time, frequency is bounded above as well. When the condition $H_a \gg H$ holds, the recognizable variation of $\mathbf{m}$ from the easy axis occurs at frequencies close to the resonance \cite{PhysRevLett.97.227202} $\Omega_{r}=\mu_{0} \gamma H_{a}$, where $\Omega_{r}$ is the resonance frequency and $\Omega_{r} \approx 2.55\cdot10^{9} \mathrm{Hz}$ for the considered nanoparticle. The last requirement to the frequency arises from the condition which represents the validity of the Stokes approximation for friction momentum\cite{Frenkel:106808}: $\mathrm{Re} = \rho_{l} \Omega_S R^{2}/ \eta \sim 10$. Here $\mathrm{Re}$ is the so-called Reynolds number, $\rho_{l}$ is the liquid density, $\Omega_S$ is the corresponding character frequency. Straightforward calculations give $\Omega_S \sim 10^{12}$ in our case. Summarizing one can obtain  $\Omega \ll \mathrm{min}[\Omega_{r}, \Omega_{S}]$ or $\Omega \ll 2.55\cdot10^{9}\mathrm{Hz}$.

Therefore, the frequency interval, where dynamical approach is valid for calculation, is $ \Omega = (10^{4}-10^{8}) \mathrm{Hz}$ that is acceptable for the magnetic fluid hyperthermia method.

\subsection{Main equations in spherical coordinates}
\label{Main_Eq_f}

Since the torque parallel to the magnetic moment is absent, two angular coordinates are necessary for the description of nanoparticle rotation. It is reasonable to use the polar and azimuthal spherical coordinates to this purpose. Let substitute the relationships $\boldsymbol{\upomega}=\mathbf{m}\times \dot{\mathbf{m}} $ and $\dot{\boldsymbol{\upomega}}=\mathbf{m}\times \ddot{\mathbf{m}}$, which follow from the second equation in (\ref{eq:Main_Eq}), into the first one. As a result, we have
\begin{equation}
    \mathbf{m}\times(\ddot{\mathbf{m}}-\dfrac{1}{\tau_0^2}\mathbf{h}+\dfrac{1}{\tau_r}\dot{\mathbf{m}})=0\\
    \label{eq:Main_Eq_Trans1}
\end{equation}
or after the corresponding transformations
\begin{equation}
    \ddot{\mathbf{m}}-\dfrac{1}{\tau_0^2}\mathbf{h}+\dfrac{1}{\tau_r}\dot{\mathbf{m}}= A(t)\mathbf{m},\\
    \label{eq:Main_Eq_Trans2}
\end{equation}
where function $A(t)$ satisfies the condition $\mathbf{m}^2=1$. Multiplying Eq.~(\ref{eq:Main_Eq_Trans2}) by $\mathbf{m}$ and taking into account the condition $\dot{\mathbf{m}}\mathbf{m}=0$, one can find
\begin{equation}
    \ddot{\mathbf{m}}+\dfrac{1}{\tau_r}\dot{\mathbf{m}} + \dfrac{1}{\tau_0^2}(\mathbf{m}\mathbf{h})\mathbf{m}+ \dot{\mathbf{m}}^2\mathbf{m}= \dfrac{1}{\tau_0^2}\mathbf{h}.\\
    \label{eq:Main_Eq_Trans3}
\end{equation}
Using the standard representation of the cartesian components of $\mathbf{m}$
in spherical coordinates $m_x = \sin \theta \cos \varphi, m_y = \sin \theta \sin \varphi, m_z = \cos \theta$, Eq.~(\ref{eq:Main_Eq_Trans3}) can be written with respect to the polar and azimuthal angles $\theta$ and $\varphi$, respectively (see Fig.~\ref{fig:model_1}):
\begin{figure}
    \centering
    \includegraphics {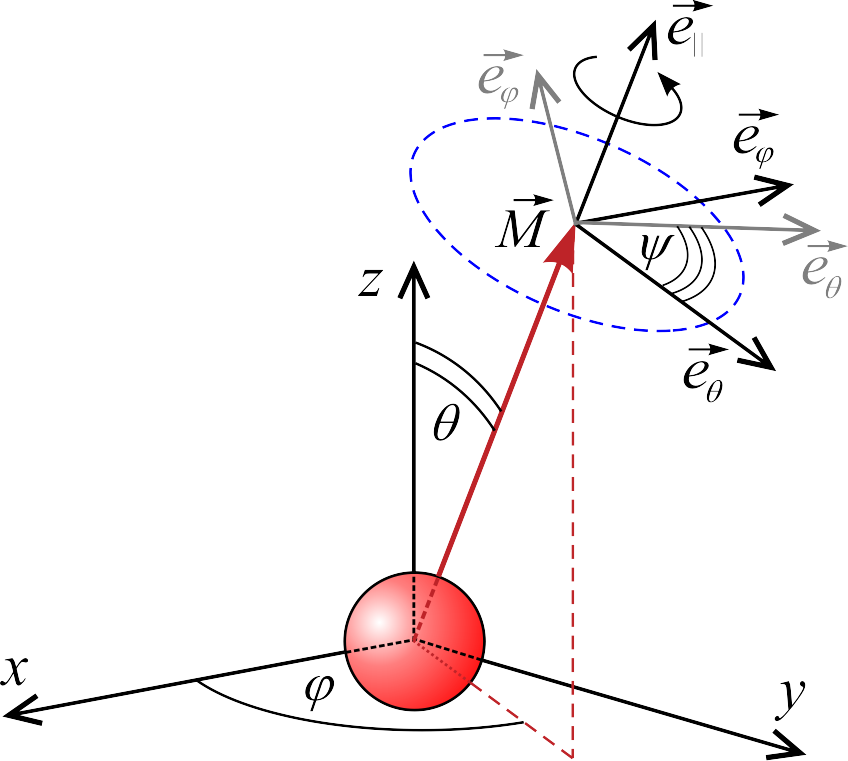}
    \caption {\label{fig:model_1} (Color online)
    Schematic representation of the model and the coordinate systems.}
\end{figure}
\begin{equation}
\begin{array}{lcl}
    \left(\ddot{\theta}+\dfrac{\dot{\theta}}{\tau_r}\right)\sin \theta  - \dot{\varphi}^2 \sin^2 \theta \cos \theta = - \dfrac{h_z + \mathbf{h}\mathbf{m}\cos\theta}{\tau_0^2},\\
    \left(\ddot{\varphi}+\dfrac{\dot{\varphi}}{\tau_r}\right)\sin \theta  + 2\dot{\varphi}\dot{\theta} \cos \theta =  \dfrac{\left(h_y \cos \varphi  - h_x \sin \varphi\right)}{\tau_0^2},\\
    \label{eq:Main_Eq_spher}
\end{array}
\end{equation}
where $h_{i} (i = x,y,z)$ are the projections of external field on the cartesian coordinates. The last equation system is irrelevant to the external field type and is convenient for finding the analytical solutions.

\subsection{Main equations: first order explicit form}
\label{Main_Eq_s}

System of Eqs.~(\ref{eq:Main_Eq_spher}) is good for the theoretical analysis, but does not fit to the well known procedure of numerical simulation. That is why one needs to rewrite Eqs.~(\ref{eq:Main_Eq}) in the explicit form which allows transformation to the difference scheme. To this end, let consider the coordinate system of unit vectors $\mathbf{e}_{\theta}$, $\mathbf{e}_{\varphi}$, $\mathbf{e}_{||}$ represented in Fig.~\ref{fig:model_1} and introduce the angle $\psi$ responsible for the rotation of the nanoparticle around the magnetic moment.

The transition from laboratory coordinate system to the new one is performed using the orthogonal transformation matrix, which can be found as the product of the rotational matrixes, which are responsible for turning about $x$ and $z$ axis respectively.
\begin{equation}
\begin{array}{lcl}
    \textbf{C} \!\!&=&\!\!
    \begin{array}{lcl}
        \left(
        \begin{array}{ccc}
            \cos\theta & 0 & - \sin\theta \\
            0 & 1 & 0 \\
            \sin\theta & 0 & \cos\theta \\
        \end{array}
        \right)
    \end{array}
    \cdot
    \begin{array}{lcl}
        \left(
        \begin{array}{ccc}
            \cos\varphi & \sin\varphi & 0 \\
            - \sin\varphi & \cos\varphi & 0\\
            0 & 0 & 1 \\
        \end{array}
    \right)
    \end{array}
    =\\ [8pt]
    \!\!&=&\!\!
    \begin{array}{lcl}
        \left(
        \begin{array}{ccc}
            \cos\theta \cos\varphi & \cos\theta \sin\varphi & - \sin\theta \\
            -\sin\varphi  & \cos\varphi & 0   \\
            \sin\theta \cos\varphi &  \sin\theta \sin\varphi & \cos\theta \\
        \end{array}
        \right),
        \label{eq:C}
    \end{array}
\end{array}
\\
\end{equation}
Correspondingly, the inverse transition can be realized by the matrix
\begin{equation}
\textbf{C}^{-1} =
  \begin{array}{lcl}
   \left(
     \begin{array}{ccc}
       \cos\theta \cos\varphi & -\sin\varphi & \sin\theta \cos\varphi \\
       \cos\theta \sin\varphi & \cos\varphi & \sin\theta \sin\varphi \\
       - \sin\theta & 0 & \cos\theta \\
     \end{array}
   \right).
   \\
   \label{eq:C_1}
\end{array}
\end{equation}
Then, following Goldstein \cite{Goldstein} let associate the time derivations of angles $\theta$, $\varphi$ and $\psi$ with the corresponding vectors
\begin{equation}
    \boldsymbol{\upomega}_{\vartheta}=\mathbf{e}_{\varphi}\dot{\theta},
    \boldsymbol{\upomega}_{\phi}=\mathbf{e}_{z}\dot{\varphi},
    \boldsymbol{\upomega}_{\psi}=\mathbf{e}_{||}\dot{\psi}.
    \label{eq:vec_w_raw}
\end{equation}
Using the orthogonal transformation obvious from Fig.~\ref{fig:model_1}, full projections of the angular velocity $\boldsymbol{\upomega} = (\omega_{\vartheta}, \omega_{\phi}, \omega_{\psi})$ on the directions of $\mathbf{e}_{\theta}$, $\mathbf{e}_{\varphi}$, $\mathbf{e}_{||}$ can be written as
\begin{equation}
\left(
  \begin{array}{c}
    {\omega}_{\theta} \\
    {\omega}_{\varphi} \\
    {\omega}_{||} \\
  \end{array}
\right)
=
\left(
  \begin{array}{c}
    - \dot{\varphi} \sin\theta  \\
    \dot{\theta} \\
    \dot{\psi} +\dot{\varphi} \cos\theta \\
  \end{array}
\right).
   \\
   \label{eq:vec_w_spher}
\end{equation}
Finally, using the well known relationship
\begin{equation}
\left(
  \begin{array}{c}
    \omega_{x} \\
    \omega_{y} \\
    \omega_{z} \\
  \end{array}
\right)
= \textbf{C}^{-1}
\left(
  \begin{array}{c}
    \omega_{\theta} \\
    \omega_{\varphi} \\
    \omega_{||} \\
  \end{array}
\right),
\\
\end{equation}
one can perform the transition to the cartesian coordinates
\begin{equation}
\left(
  \begin{array}{c}
    \omega_{x} \\
    \omega_{y} \\
    \omega_{z} \\
  \end{array}
\right)
=
\left(
  \begin{array}{c}
    -\dot{\theta}\sin\varphi + \dot{\psi}\sin\theta\cos\varphi  \\
    \dot{\theta}\cos\varphi + \dot{\psi}\sin\theta\sin\varphi \\
    \dot{\varphi} + \dot{\psi}\cos\theta \\
  \end{array}
\right).
   \\
   \label{eq:vec_w_cart}
\end{equation}
Here it is important to underline that the latter equations are similar (but not the same) to the corresponding equations for the Euler angles \cite{Goldstein}, because of the way they were obtained. From Eqs.~(\ref{eq:vec_w_cart}) the first order differential equations in explicit form with respect to the polar and azimuthal angles of the nanoparticle are derived
\begin{equation}
\begin{array}{lcl}
    \dot{\theta} = \omega_{y} \cos\varphi - \omega_{x} \sin\varphi, \\
    \dot{\varphi} = \omega_{z} - \left(\omega_{x} \cos\varphi + \omega_{y} \sin\varphi\right) \cot \theta. \\
    \\
   \label{eq:theta_phi}
\end{array}
\end{equation}
The next three equations can be extracted directly from Eqs.~(\ref{eq:Main_Eq}) using the standard representation of the vector $\mathbf{m}$ in spherical coordinates. And, as a result, one can write the main equations in the reduced form
\begin{equation}
\begin{array}{lcl}
    \dfrac{d\theta}{d\tilde{t}} = \widetilde{\omega}_y \cos\varphi - \widetilde{\omega}_x \sin\varphi, \\ [6pt]
    \dfrac{d\varphi}{d\tilde{t}} = \widetilde{\omega}_z - \left(\widetilde{\omega}_x \cos\varphi + \widetilde{\omega}_y \sin\varphi\right) \cot \theta, \\ [6pt]
    \dfrac{d\widetilde{\omega}_x} {d\tilde{t}} = \dfrac{1}{\alpha} \left( h_z\sin\theta \sin\varphi -h_y\cos\theta - \widetilde{\omega}_x \right), \\ [6pt]
    \dfrac{d\widetilde{\omega}_y} {d\tilde{t}} =  \dfrac{1}{\alpha} \left( h_x\cos\theta - h_z\sin\theta \cos\varphi + \widetilde{\omega}_y \right),\\ [6pt]
    \dfrac{d\widetilde{\omega}_z} {d\tilde{t}} =  \dfrac{1}{\alpha} \left( h_y\sin\theta \cos\varphi -h_x\sin\theta \sin\varphi  -  \widetilde{\omega}_z \right). \\ [6pt]
    \\
   \label{eq:Main_Eq_num_red}
\end{array}
\end{equation}
where $\widetilde{\omega}_i = \omega_i/\Omega_{cr},  \alpha=\tau_r^2/\tau_0^2, i = x,y,z$. These equations can be easily implemented into the numerical solution technique.

\section{ANALYTICAL RESULTS}
\label{Res}

\subsection{Circularly polarized magnetic field}
\label{An_Circ}

It is assumed that the nanoparticle is governed by external field of the following type:
\begin{equation}
    \mathbf{h}(t) = \mathbf{e}_{x} h\cos(\Omega t) + \mathbf{e}_{y}\sigma h\sin(\Omega t) + \mathbf{e}_{z} h_{z0},
    \label{eq:def_h_cp}
\end{equation}
where $h$ and $\Omega$ are the rotating field amplitude and frequency, respectively, $h_{z0}$ is the time-independent or static filed, $\sigma = \pm1$ is the identifier of the direction of rotation. In order to describe the trajectory of the nanoparticle in this case, one should substitute expression (\ref{eq:def_h_cp}) into Eqs.~(\ref{eq:Main_Eq_spher}). Using the reduced values of frequency $\widetilde{\Omega} = \Omega/\Omega_{cr}$, we obtain
\begin{equation}
\begin{array}{lcl}
    \dfrac{d^2\theta} {d^2\tilde{t}}\!+\!\dfrac{1}{\alpha}\dfrac{d\theta} {d\tilde{t}}\!-\!\left(\dfrac{d\varphi} {d\tilde{t}}\right)^2 \sin \theta \cos \theta \!=\! \dfrac{h}{\alpha} \cos\phi\cos \theta - \dfrac{h_{z0}}{\alpha}\sin \theta,\\
    \\ [6pt]
    \left(\dfrac{d^2\varphi} {d^2\tilde{t}}+\dfrac{1}{\alpha}\dfrac{d\varphi} {d\tilde{t}}\right)\sin\theta + 2 \dfrac{d\theta} {d\tilde{t}}\dfrac{d\varphi} {d\tilde{t}} \cos\theta = - \dfrac{h}{\alpha} \sin\phi,\\
    \label{eq:Main_Eq_spher_cp}
\end{array}
\end{equation}
where $\phi = \varphi-\rho \widetilde{\Omega} \tilde{t}$. The uniform rotational mode with the field frequency is the natural solution of Eqs.~(\ref{eq:Main_Eq_spher_cp}) This mode is characterized by the constant lag angle $\Phi = \lim_{\tilde{t} \to \infty} \phi$, and constant angle of the precessional cone $\Theta = \lim_{\tilde{t} \to \infty} \theta$. Substituting the latter solutions into Eqs.~(\ref{eq:Main_Eq_spher_cp}) we derive the systems of algebraic equations for the calculation of $\Phi$ and $\Theta$
\begin{equation}
\begin{array}{lcl}
    \cos \Theta\left(\widetilde{\Omega}^2\alpha\sin\Theta + h\cos\Phi \right) = h_{z0}\sin\Theta, \\
    \widetilde{\Omega}\sin\Theta =h\sin\Phi.\\
    \label{eq:Main_Eq_spher_cp_sol}
\end{array}
\end{equation}
The last equations system can be solved numerically for defined system parameters. The average value of the power loss for the uniform mode can be found using (\ref{eq:Main_Eq_spher_cp_sol}) and (\ref{eq:def_Q_red_1}) as follows
\begin{equation}
    \widetilde{Q} = \widetilde{\Omega}^{2} \sin^{2}\Theta.
    \label{eq:Q_circ}
\end{equation}
Two remarks are relevant here. Firstly, (\ref{eq:Q_circ}) for a small angle of the precession cone coincides with the results obtained by Xi \cite{0022-3727-39-22-002} in the linear approximation. And, secondly, when the static field is absent ($h_{z0} = 0$), the obtained relationships reduced to the simpler forms: $\Theta =\pi/2$, $\sin \Phi = \widetilde{\Omega}/h$, and $\widetilde{Q} = \widetilde{\Omega}^{2}$.

The solutions obtained from Eqs.~(\ref{eq:Main_Eq_spher_cp_sol}) are stable while $h_{z0} \neq 0$. In the opposite case, the stable precession is possible when $\widetilde{\Omega} \leq\ h$. In turn, if $\widetilde{\Omega} > h$ the nonuniform precession mode takes place that will be described numerically below. However, we should notice the following. When the condition $\theta =\pi/2$ holds and using the representation $\varphi(\tilde{t}) = \widetilde{\Omega}\tilde{t}-\beta(\tilde{t})$, one can write
\begin{equation}
    \dfrac{d^2\beta} {d^2\tilde{t}}+\dfrac{1}{\alpha}\dfrac{d\beta} {d\tilde{t}} +
    \dfrac{h}{\alpha}\sin\beta = 0.\\
    \label{eq:Main_Eq_pend}
\end{equation}
This is the well known equation for a damped pendulum under constant torque \cite{doi:10.1142/7861}.

\subsection{Linearly polarized magnetic field}
\label{An_Lin}

In order to simplify the further analysis, let consider external field of the type
\begin{equation}
    \mathbf{h}(t) = \mathbf{e}_{z} h\cos(\widetilde{\Omega} \tilde{t}).
    \label{eq:def_h_lp}
\end{equation}
After substitution of (\ref{eq:def_h_lp}) into Eqs.~(\ref{eq:Main_Eq_spher}) the equations of motion take the following form:
\begin{equation}
\begin{array}{lcl}
    \dfrac{d^2\theta} {d^2\tilde{t}}+\dfrac{1}{\alpha}\dfrac{d\theta} {d\tilde{t}}  - \left(\dfrac{d\varphi} {d\tilde{t}}\right)^2 \sin \theta \cos \theta = - \dfrac{h}{\alpha}\sin \theta  \cos(\widetilde{\Omega}\tilde{t}), \\
    \left(\dfrac{d^2\varphi} {d^2\tilde{t}}+\dfrac{1}{\alpha}\dfrac{d\varphi} {d\tilde{t}}\right)\sin\theta + 2 \dfrac{d\theta} {d\tilde{t}}\dfrac{d\varphi} {d\tilde{t}} \cos\theta = 0.\\
    \label{eq:Main_Eq_spher_lp}
\end{array}
\end{equation}
From the second equation of Eqs.~(\ref{eq:Main_Eq_spher_lp}) one can obtain
\begin{equation}
    \left(\dfrac{d\varphi} {d\tilde{t}}\right)^{-1} \dfrac{d^2\varphi} {d^2\tilde{t}}+\dfrac{1}{\alpha}  = -2 \dfrac{d\theta} {d\tilde{t}} \dfrac{\cos\theta} {\sin\theta}.
    \label{eq:Main_Eq_spher_lp1}
\end{equation}
The latter can be easily integrated
\begin{equation}
    \left|\dfrac{d\varphi}{d\tilde{t}}\right| = \left|\dfrac{d\varphi(0)}{d\tilde{t}}\right| \sin^{-2}\theta\exp\left( -\dfrac{\tilde{t}}{\alpha} \right).
    \label{eq:Main_Eq_spher_lp2}
\end{equation}
Due to the exponential decaying, the long time evolution of the azimuthal coordinate can be neglected and the motion should be considered as planar. Therefore, the first expression in Eqs.~(\ref{eq:Main_Eq_spher_lp}) yields
\begin{equation}
\dfrac{d^2\theta} {d^2\tilde{t}}+\dfrac{1}{\alpha}\dfrac{d\theta} {d\tilde{t}}  = - \dfrac{h}{\alpha}\sin \theta  \cos(\widetilde{\Omega}\tilde{t}). \\
\end{equation}
This equation has the following solution in the approximation of small inertia momentum ($I \rightarrow 0$, $\ddot{\theta}\equiv0$)
\begin{equation}
    \tan (\theta/2) = \tan (\theta_0/2)\exp\left[-\sin(\widetilde{\Omega}\tilde{t})\dfrac{h}{\widetilde{\Omega}}\right],\\
    \label{eq:Main_Eq_lp_without_I}
\end{equation}
where $\theta_0$ is the initial polar angle of $\mathbf{m}$. Notice here that the similar equation was reported in \cite{doi:10.1063/1.4737126}, but for the description of the joint internal magnetic dynamics and mechanical rotation in linear approximation. One needs to note that there it has been obtained within controversial basic equations.

Using (\ref{eq:def_Q_red_1}), the average value of the power loss for the uniform precession mode can be written as
\begin{equation}
    \widetilde{Q} = \dfrac{\widetilde{\Omega}^{2}h}{2\pi} \int_{0}^{\frac{\widetilde{\Omega}} {2\pi}} d\tilde{t} \tanh \left[ \dfrac{h}{\widetilde{\Omega}}\sin(\widetilde{\Omega} \tilde{t}) - \dfrac{x_0}{2}\right] \sin (\widetilde{\Omega} \tilde{t}),
    \label{eq:Q_lin}
\end{equation}
where $x_0 = \ln \left(\tan^2 \theta_0/2\right)$ is the constant defined by the initial state of $\mathbf{m}$.

\subsection{High frequency limit: linear approximation}
\label{An_Hf}
\begin{figure}
    \centering
    \includegraphics {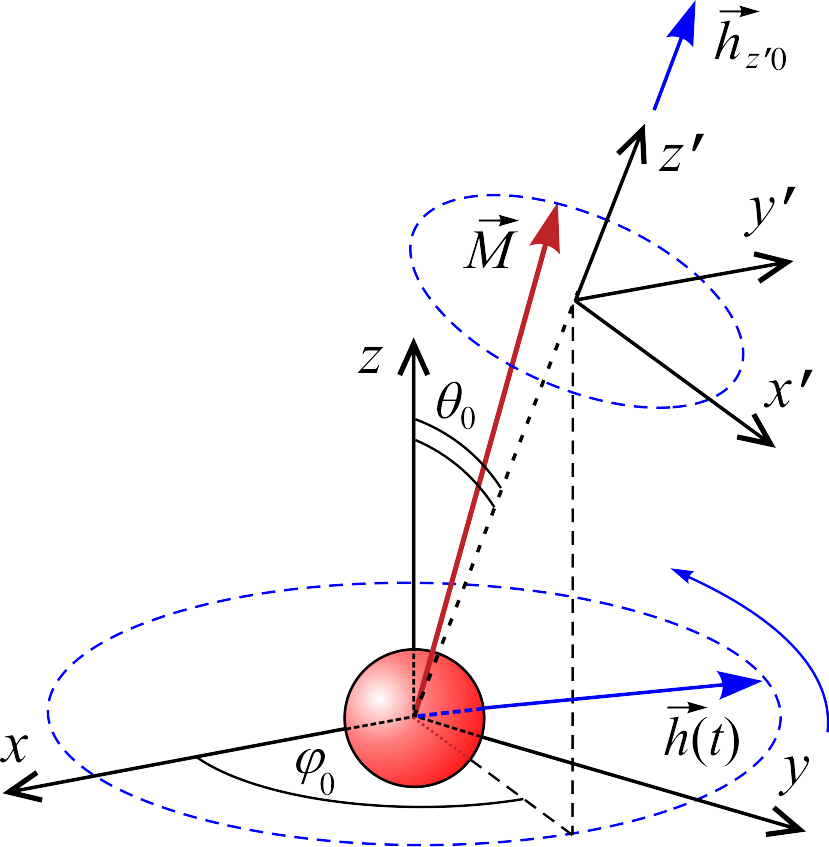}
    \caption {\label{fig:model_2} (Color online)
    Schematic representation of the model and the rotated coordinate system for the high frequency limit case.}
\end{figure}

When the condition $\widetilde{\Omega} \gg h$ holds, the magnetic moment performs small oscillations. Equilibrium point for these oscillations is conditioned either by the initial position of the magnetic moment or by the direction of the additional static field $\mathbf{h}_{z'0}$. To simplify the further analysis we assume that the direction of $\mathbf{h}_{z'0}$ as well as the initial direction of $\mathbf{m}$ are defined by the angles $\theta_0$ and $\varphi_0$ (see Fig.~\ref{fig:model_2}). The static field is non-perpendicular to the alternating field that expands the limits of the used model. Among others, the local dipole field, which is induced by all the particles in a ferrofluid and acts on the given particle, can be considered as the filed $\mathbf{h}_{z'0}$.

In Fig.~\ref{fig:model_2} the new coordinate system defined by unit vectors ($\mathbf{e}_{x'}, \mathbf{e}_{y'}, \mathbf{e}_{z'}$) is depicted. This system is fixed, but turned with respect to the laboratory one by the angles $\theta_0$ and $\varphi_0$. The alternating field (\ref{eq:def_h_cp}) in these coordinates can be transformed using the rotation matrix similar to (\ref{eq:C}), but written for the angles $\theta_0$ and $\varphi_0$ instead of $\theta$ and $\varphi$ correspondingly. Taking into account that the static field is collinear to $\mathbf{e}_{z'}$ ($\mathbf{h}_{z'0} = \mathbf{e}_{z'}h_{z'0}$), the total external field can be written as

\begin{figure}
    \centering
    \includegraphics {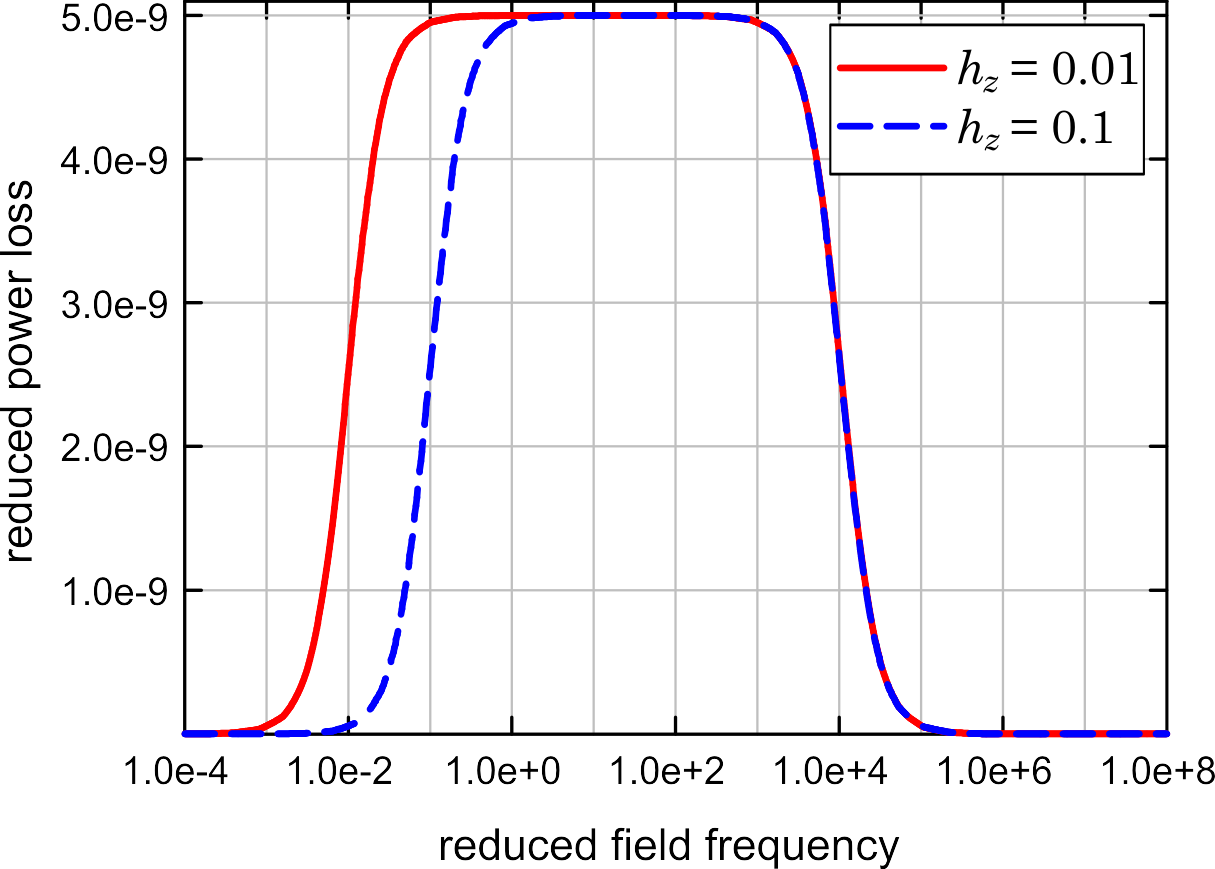}
    \caption {\label{fig:Q_hf} (Color online)
    Results: The power loss in the high frequency approximation according to (\ref{eq:Q_hf}). Circularly polarized field  (\ref{eq:def_h_cp}) is acting. The values of the system parameters are chosen as $R=20\mathrm{nm}$, $\mathcal{M} = 3.38\cdot10^{5} \mathrm{A/m}$, $\eta = 5\cdot10^{-3} \mathrm{Pa}\cdot\mathrm{s}^{1}$, $\sigma = 1$, $h=0.05$, $\theta_0 = 0.01$. The possibility of control of $\widetilde{Q}(\widetilde{\Omega})$ by the value of $h_{z'0}$ is concluded.}
\end{figure}

\begin{equation}
\mathbf{h}' = \left(
  \begin{array}{c}
    h \cos\theta_0 \cos \varphi_0 \cos (\widetilde{\Omega} \tilde{t}) +  \sigma h \cos\theta_0 \sin\varphi_0 \sin(\widetilde{\Omega} \tilde{t})\\
    -h \sin \varphi_0 \cos(\widetilde{\Omega} \tilde{t}) +  \sigma h \cos\varphi_0 \sin(\widetilde{\Omega} \tilde{t})\\
    h \sin\theta_0 [\cos \varphi_0 \cos(\widetilde{\Omega} \tilde{t}) + \sigma \sin \varphi_0 \sin(\widetilde{\Omega} \tilde{t})] + h_{z'0}\\
  \end{array}
\right).
   \\
   \label{eq:h_dash}
\end{equation}
While $\sigma = \pm1$, (\ref{eq:h_dash}) represents the circularly polarized field and condition $\sigma = 0$ defines the linearly polarized one. For the  description of small oscillations let express vector of the magnetic moment in the new coordinate system as $\mathbf{m}'=\mathbf{e}_{x'}m_{x'} +\mathbf{e}_{y'}m_{y'} + \mathbf{e}_{z'}$. Within the assumption $m_{x'},m_{y'},\omega_{x'}, \omega_{y'}, \omega_{z'} \sim h$ and with the first-order accuracy, the system (\ref{eq:Main_Eq}) takes the form
\begin{equation}
\begin{array}{lcl}
    \dfrac{d m_{x'}} {d\tilde{t}} \!\!&=&\!\! \widetilde{\omega}_{y'}, \\ [6pt]
    \dfrac{d m_{y'}} {d\tilde{t}} \!\!&=&\!\!- \widetilde{\omega}_{x'},\\ [6pt]
    \dfrac{d {\widetilde{\omega}}_{x'}} {d\tilde{t}} \!\!&=&\!\! \dfrac{1}{\alpha}[m_{y'}h_{z'0} + h \sin \varphi_0 \cos(\widetilde{\Omega} \tilde{t}) - \\
    &&- \sigma h \cos\varphi_0 \sin(\widetilde{\Omega} \tilde{t}) - \widetilde{\omega}_{x'}],\\ [6pt]
    \dfrac{d {\widetilde{\omega}}_{y'}} {d\tilde{t}} \!\!&=&\!\! \dfrac{1}{\alpha} [- m_{x'}h_{z'0} +  h \cos\theta_0 \cos \varphi_0 \cos (\widetilde{\Omega} \tilde{t}) + \\
     &&+ \sigma h \cos\theta_0 \sin\varphi_0 \sin(\widetilde{\Omega} \tilde{t}) - \widetilde{\omega}_{y'}],\\ [6pt]
    \dfrac{d {\widetilde{\omega}}_{z'}} {d\tilde{t}} \!\!&=&\!\! - \dfrac{1}{\alpha} \widetilde{\omega}_{z'}.\\
    \\
    \label{eq:Main_Eq_linear}
\end{array}
\end{equation}
We find the solution of this system of linear equations in the standard form
\begin{equation}
\begin{array}{lcl}
    m_{x'} = a\cos(\widetilde{\Omega} \tilde{t}) + b\sin(\widetilde{\Omega} \tilde{t}),\\
    m_{y'} = c\cos(\widetilde{\Omega} \tilde{t}) + d\sin(\widetilde{\Omega} \tilde{t}),\\
    \widetilde{\omega}_{x'} = f\cos(\widetilde{\Omega} \tilde{t}) + g\sin(\widetilde{\Omega} \tilde{t}),\\
    \widetilde{\omega}_{y'} = k\cos(\widetilde{\Omega} \tilde{t}) + l\sin(\widetilde{\Omega} \tilde{t}),\\
    \widetilde{\omega}_{z'} = p\cos(\widetilde{\Omega} \tilde{t}) + q\sin(\widetilde{\Omega} \tilde{t}).\\
    \label{eq:Main_Eq_linear_sol}
\end{array}
\end{equation}
Substituting (\ref{eq:Main_Eq_linear_sol}) into (\ref{eq:Main_Eq_linear}) and using the linear independence of the trigonometric functions $\sin(\widetilde{\Omega} \tilde{t})$ and $\cos(\widetilde{\Omega} \tilde{t})$, one straightforwardly obtains the constants which define $\mathbf{m}$
\begin{equation}
\begin{array}{lcl}
    \displaystyle a = - h \cos\theta_0 \frac{(\alpha{\widetilde{\Omega}}^2 - h_{z'0}) \cos\varphi_0 + \sigma \widetilde{\Omega} \sin\varphi_0} {{\widetilde{\Omega}}^2 + (\alpha {\widetilde{\Omega}}^2- h_{z'0})^2},\\
    \displaystyle b =  - h \cos\theta_0 \frac{\sigma(\alpha{\widetilde{\Omega}}^2 - h_{z'0}) \sin\varphi_0 - \widetilde{\Omega} \cos\varphi_0} {{\widetilde{\Omega}}^2 + (\alpha {\widetilde{\Omega}}^2- h_{z'0})^2},\\
    \displaystyle c = h \frac{(\alpha{\widetilde{\Omega}}^2 - h_{z'0}) \sin\varphi_0 - \sigma \widetilde{\Omega}\cos\varphi_0} {{\widetilde{\Omega}}^2 + (\alpha {\widetilde{\Omega}}^2- h_{z'0})^2},\\
    \displaystyle d =  - h \frac{\sigma(\alpha{\widetilde{\Omega}}^2 - h_{z'0}) \cos\varphi_0 + \widetilde{\Omega} \sin\varphi_0} {{\widetilde{\Omega}}^2 + (\alpha {\widetilde{\Omega}}^2- h_{z'0})^2}.\\
    \\
    \label{eq:Main_Eq_linear_koef}
\end{array}
\end{equation}
Finally, after substitution of (\ref{eq:Main_Eq_linear_sol}), (\ref{eq:Main_Eq_linear_koef}) into Eq.~(\ref{eq:def_Q_red_1}) and further integration of the latter, the relationship for the power loss can be written as
\begin{equation}
    \widetilde{Q} = \dfrac{\widetilde{\Omega}^{2}h^2 D}{2 [\widetilde{\Omega}^2 + (\alpha {\widetilde{\Omega}}^2- h_{z'0})^2]},
    \label{eq:Q_hf}
\end{equation}
where
\begin{equation}
D =  \cos^2 \theta_0(\cos^2 \varphi_0 + \sigma^2\sin^2 \varphi_0) + \sigma^2\cos^2 \varphi_0 + \sin^2 \varphi_0
\label{eq:D}
\end{equation}
is the factor, defined by the initial state of the magnetic moment. Let analyze (\ref{eq:Q_hf}). 1) If the circularly polarized field is acting, the factor $D = 1 + \cos^2 \theta_0$ ($1 \leq D \leq 2$). At the same time, for linearly polarized one it is $D = \cos^2 \varphi_0 (\cos^2 \theta_0 - 1) + 1$ ($0 \leq D \leq 1$). Therefore, the circular polarization can be, at least in two times more efficient for heating in comparison with liner one while  $\widetilde{\Omega} \gg h$. 2) When $h_{z'0} = 0$, the oscillations occur around the initial state, and the power loss depends on the polar angle $\theta_0$ of the initial state. 3) When the condition $h_{z'0} \gg  \widetilde{\Omega}$ holds, the power loss depends on the static field value, the alternating field frequency and amplitude: the asymptotic have the form $\widetilde{Q} \sim \widetilde{\Omega}^2 h^2/h_{z'0}^2$. 4) Otherwise, for small static fields, $h_{z'0} \ll  \widetilde{\Omega}$, the asymptotic is the following $\widetilde{Q} \sim h^2$ that means the power loss is determined only by the alternating field amplitude. This fact gives the opportunity to control the heating rate by the static field $h_{z'0}$ in the intermediate frequency range (See Fig.~\ref{fig:Q_hf}). 5) And, finally, if the particle is small enough or frequency is too large ($\widetilde{\Omega} \gg 1/\alpha$), then the asymptotic is $\widetilde{Q} \sim h^2/(\alpha^2 \widetilde{\Omega}^4 )$. Therefore, the effects caused by the inertia momentum of the particles take place, when $\widetilde{\Omega} > 1/\alpha$ or the critical value $\Omega_I = \Omega_{cr}/\alpha$ exists ($\Omega_I \approx 3.83\cdot10^{10} \mathrm{Hz}$ for the considered nanoparticle of maghemite $R=20\mathrm{nm}$ \cite{C3RA45457F}). These estimations are close to analogous in \cite{Raikher_1994}. Thus, while the condition $\Omega \ll \Omega_I$ holds, the inertia term can be neglected and the equations of motion Eqs.~(\ref{eq:Main_Eq}) are reduced to the widely used form

\begin{equation}
    \dot{\mathbf{m}}=\dfrac{\tau_r}{\tau_0^2}\boldsymbol{\upomega}\times(\mathbf{m}\times\mathbf{h}), \\
    \label{eq:Eq_Reduced}
\end{equation}
or in the spherical coordinates
\begin{equation}
\begin{array}{lcl}
    \dot{\theta} =  \cos\theta(h_x\cos\varphi + h_y\sin\varphi) - h_{z0} \sin\theta, \\ [6pt]
    \dot{\varphi} =  \dfrac{1}{\sin\theta}(h_y\cos\varphi - h_x\sin\varphi).  \\
    \label{eq:Eq_Reduced_Spher}
\end{array}
\end{equation}

\section{NUMERICAL RESULTS: GENERAL CASE}

\label{An_Num}
\begin{figure}
    \centering
    \includegraphics {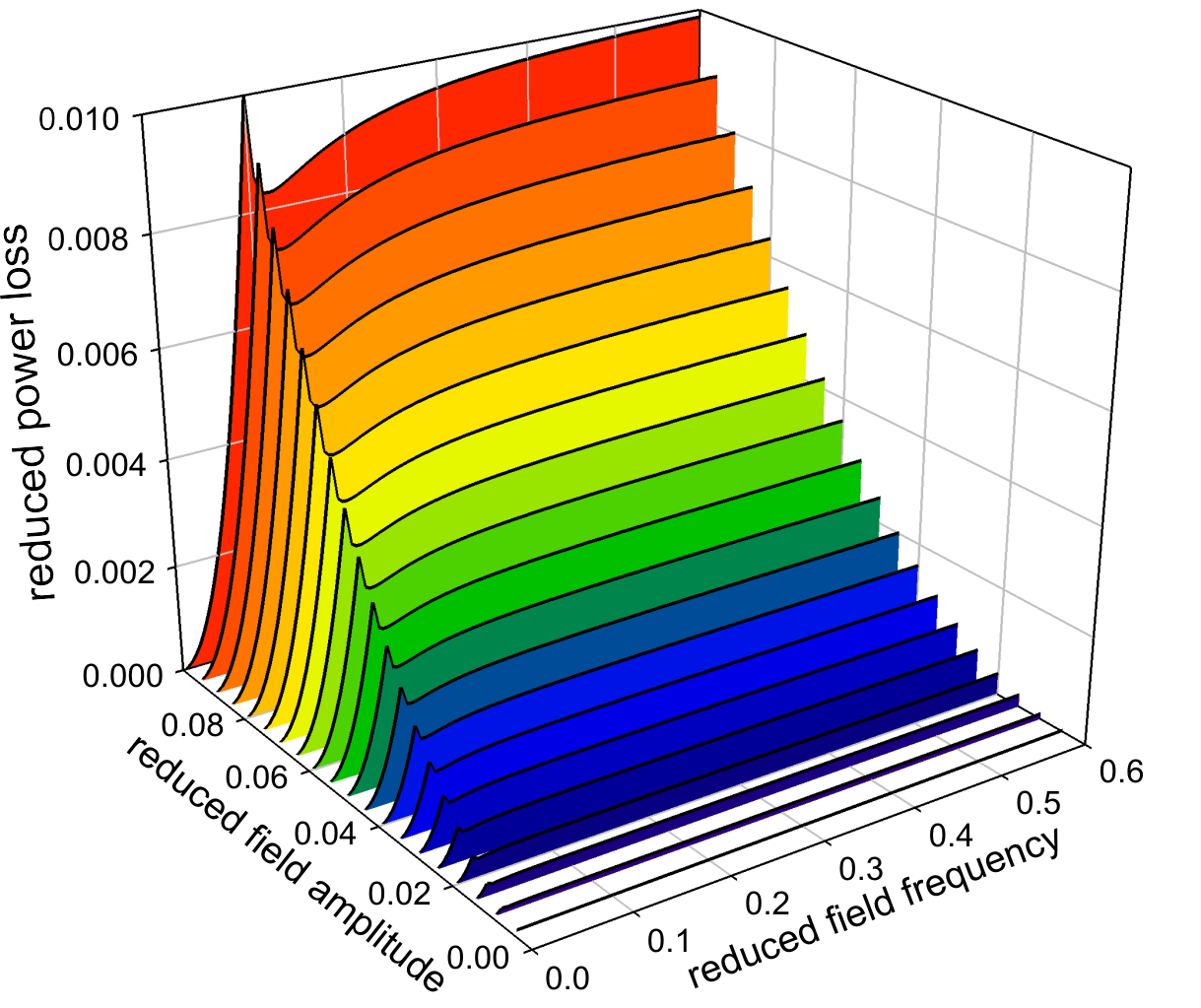}
    \includegraphics {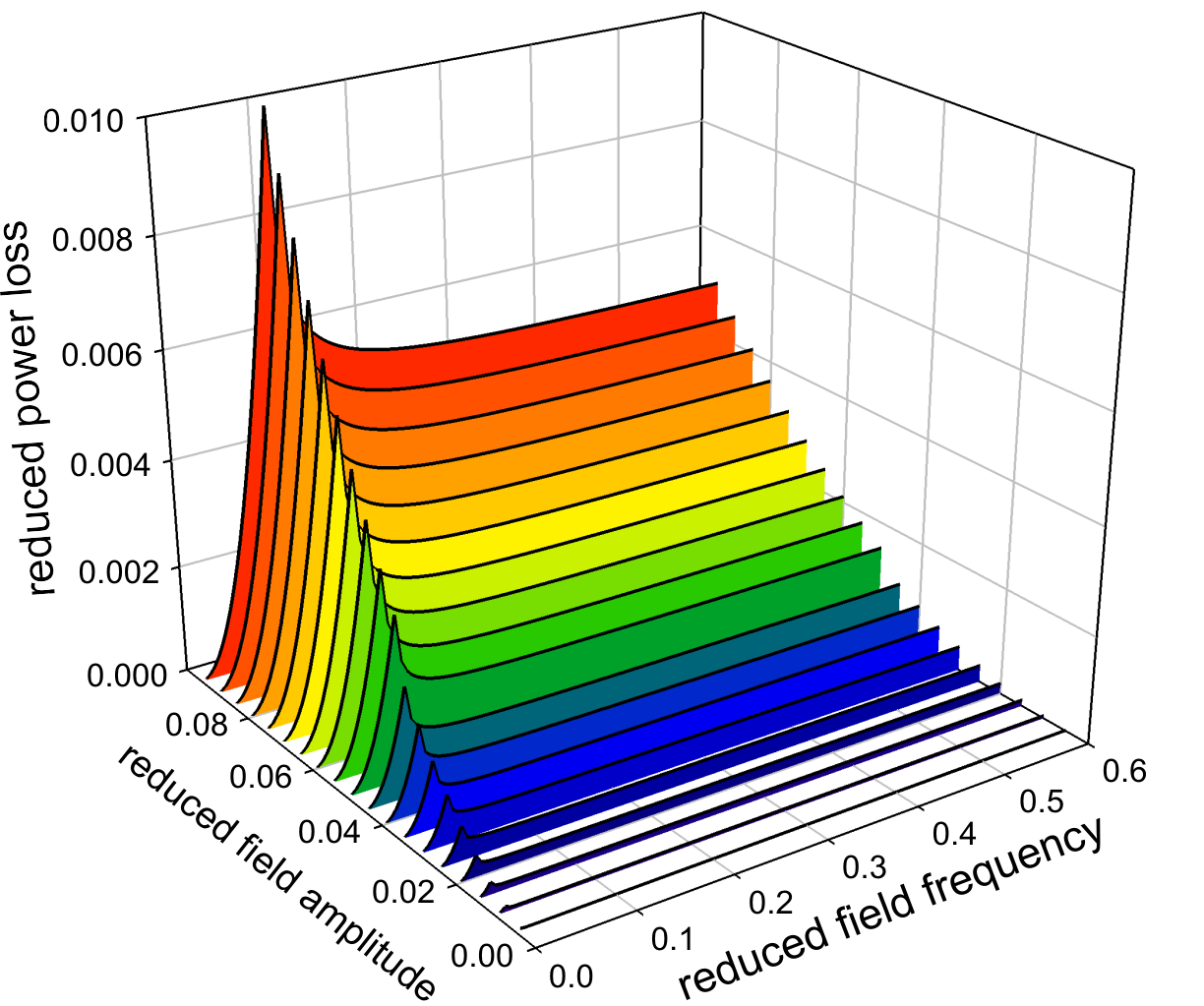}
    \caption {\label{fig:Res_cp} (Color online)
    The simulation results for the power loss obtained for different initial conditions of the magnetic moment: $\theta_0 = 0.01$ (a), $\theta_0 = \pi/2-0.01$ (b). Circularly polarized field (\ref{eq:def_h_cp}) is acting. The values of the system parameters are chosen as $R=20\mathrm{nm}$, $\mathcal{M} = 3.38\cdot10^{5} \mathrm{A/m}$, $\eta = 5\cdot10^{-3} \mathrm{Pa}\cdot\mathrm{s}^{1}$, $\sigma = 1$. The saturated character of the frequency dependencies and the reduction of the power loss due to the nonuniform mode generation are observed.}

\end{figure}

To validate the obtained above analytical results, describe the power loss behavior in the whole range of parameters, and finally, visualize the data, the numerical simulation is demanded. The latter was performed using Eqs.~(\ref{eq:Main_Eq_num_red}) and the Runge-Kutta method (the fourth-order method). Analyze first the case of the circularly polarized field action, see (\ref{eq:def_h_cp}), $\sigma = 1$. As follows from the analytical results discussed above, when $h > \widetilde{\Omega}$, particle is rotated uniformly and all contributions into the power loss are due to this rotation. The rotation is realized with the field frequency, and the losses are proportional to the particle kinetic energy. Therefore, $\widetilde{Q} \sim \widetilde{\Omega}^2$б and it was confirmed numerically. When $h \ll \widetilde{\Omega}$, oscillations occur with the field frequency, but amplitude decreases when the frequency grows. In this case $\widetilde{Q}$ almost not increase with $\widetilde{\Omega}$ in the oscillations mode. These facts are confirmed numerically. At the same time, when $h < \widetilde{\Omega}$ and $h \sim \widetilde{\Omega}$, the dynamics become more complicated. This is expressed in the generation of the nonuniform mode similarly to the case of the magnetic moment dynamics inside the particle \cite{PhysRevLett.86.724, 0953-8984-21-39-396002}. In this mode the polar angle $\theta$ of the magnetic moment is changed periodically in time with a period, which does not coincide with the field's one. The similar oscillations are demonstrated by the azimuthal angle $\varphi$ together with linear growth in time. This type of motion is characterized by smaller instantaneous angular velocity of the nanoparticle that leads to the decrease in the power loss. It is expressed in a considerable drop of $\widetilde{Q}(\widetilde{\Omega})$ for the fixed amplitude $h$. The results of the simulations set as 3D waterfall plots are shown in Fig.~\ref{fig:Res_cp}.

In contrary with the case when the nanoparticle is fixed rigidly in a solid matrix, here the dependence $\widetilde{Q}(\widetilde{\Omega})$ does not exhibit a resonant behaviour and have non-zero high-frequency asymptotic. The explanation is could be the following. While the field frequency grows, the average angular velocity tends to zero, the oscillation frequency tends to $\widetilde{\Omega}$, and the oscillation amplitude tends to the values, predicted by (\ref{eq:Main_Eq_linear_koef}). Therefore, $\widetilde{Q}$ tends to the value predicted by (\ref{eq:Q_hf}) fast enough with $\widetilde{\Omega}$ (see Fig.~\ref{fig:Res_cp}).

Another feature of the nonuniform mode consists in the dependence of trajectory on the initial conditions of $\mathbf{m}$. This fact is caused by the absence of any local minimums which can capture the magnetic moment when the external field is removed. Such difference leads to the dependence of the power loss $\widetilde{Q}$ on $\theta_0$, similarly to the case of the rotational oscillation limit, and its asymptotic values are defined by expression (\ref{eq:Q_hf}). Here and further two limit situations $\theta_0 \rightarrow \pi/2$, and $\theta_0 \rightarrow 0$ are considered. Special attention deserves the former case. Here the validity of Eq.~(\ref{eq:Main_Eq_pend}) is confirmed numerically: only $\varphi$ angle is changed in time, while $\theta$ remains unchangeable and equal to $\theta_0 = \pi/2$.

\begin{figure}
    \centering
    \includegraphics {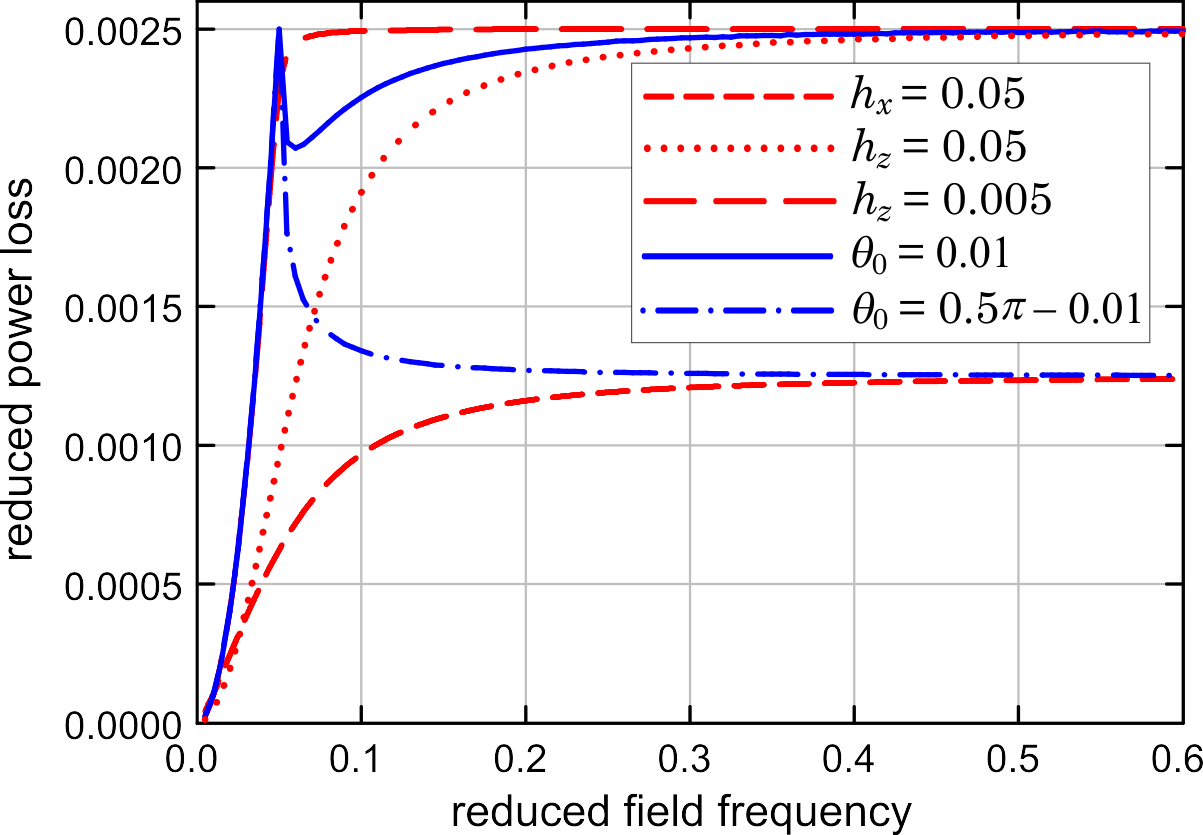}
    \caption {\label{fig:res_cp_h} (Color online)
    The comparison of the numerical results for the power loss obtained for different initial conditions and additional static fields. Circularly polarized field (\ref{eq:def_h_cp}) is acting. The values of the system parameters are chosen as $R=20\mathrm{nm}$, $\mathcal{M} = 3.38\cdot10^{5} \mathrm{A/m}$, $\eta = 5\cdot10^{-3} \mathrm{Pa}\cdot\mathrm{s}^{1}$, $h = 0.05$, $\sigma = 1$. The curves obtained for stable modes in the absence of a static field are shown in blue color. The curves obtained after the corresponding static field application (which are non-sensitive to the initial conditions) are shown in red color. The possibility of control of the heating rate by a static field is suggested.}
\end{figure}

The influence of the static field on the magnetic moment dynamics consists, firstly, in a partial pinning of $\mathbf{m}$ along of the static field $\mathbf{h}_0$ direction. This pinning leads to the decrease in the magnetic moment mobility and its response on the alternating field. Obviously, changes in $\widetilde{Q}$ value will be determined by the direction of the static field with respect to the alternating one. It needs to note that since the static field establishes a new attraction point for $\mathbf{m}$, the trajectory dependence on the initial position of the magnetic moment, when $h < \widetilde{\Omega}$, is broken. The foregoing give opportunities to reduce the power loss as well as its enhancing by the static field. Let consider the most simple and demonstrative situations: the static field lays in the polarization plane ($\mathbf{h}_0 = (h_{x0},0,0)$) and perpendicular to it ($\mathbf{h}_0 = (0, 0, h_{z0})$). Also we suppose that before $\mathbf{h}_0$ application, the stable modes of $\mathbf{m}$ are established for the initial conditions $\theta_0 \rightarrow \pi/2$ or $\theta_0 \rightarrow 0$. Thus, the following cases are possible (See Fig.~\ref{fig:res_cp_h}).

1) $\theta_0 \rightarrow \pi/2$, $\mathbf{h}_0 = (0, 0, h_{z0})$. If $h_{z0} \ll h$, the static field influences weakly on $\widetilde{Q}$ while $\mathbf{m}$ performs the uniform rotation (the condition $\widetilde{\Omega} < h$ holds). At the same time, due to the suppression of the nonuniform rotation for larger frequencies ($\widetilde{\Omega} > h$), field $\mathbf{h}_0 = (0, 0, h_{z0})$ leads to the growth of $\widetilde{Q}$ (see the blue solid and red long dashed lines in Fig.~\ref{fig:res_cp_h}). This is a remarkable effect, but it is broken for large enough ($h_{z0} \sim h$) static fields due to  $\mathbf{m}$ pinning (see the blue solid and red dotted lines in Fig.~\ref{fig:res_cp_h}).

2) $\theta_0 \rightarrow \pi/2$, $\mathbf{h}_0 = (h_{x0},0,0)$. If $h_{x0} < h$, the static field does not affect essentially on the power loss, because static field accelerates $\mathbf{m}$ while $\mathbf{h}_{0}$ and $\mathbf{h}(t)$ are codirectional, and slows $\mathbf{m}$ in the opposite case. Finally, these two effects cancel each other. But if  $h_{x0} \geq h$, $\mathbf{m}$ always performs the oscillations around  $\mathbf{h}_{0}$ direction in the $xoy$ plane instead of the rotation that leads to decrease in $\widetilde{Q}$ for all frequencies: see the blue dashed-dotted and red short dashed lines in Fig.~\ref{fig:res_cp_h}.

3) $\theta_0 \rightarrow 0$, $\mathbf{h}_0 = (h_{x0},0,0)$. The main effect of $\mathbf{h}_0 = (h_{x0},0,0)$  is the reduction of the degree of freedom numbers of $\mathbf{m}$. While $h_{x0}$ is applied, only azimuthal angle $\varphi$ is changed that causes the decrease in the nanoparticle angular velocity and the power loss consequently. This is especially expressed for $h_{x0} > h$ in comparison with the case without static field, when $\mathbf{m}$ is initially directed perpendicularly to the field polarization plane. One can observe that the corresponding dependences $\widetilde{Q}(\widetilde{\Omega})$ can be different in several times for $\widetilde{\Omega} \sim h$ and in two times for big frequencies ($\widetilde{\Omega} \gg h$), see the blue solid and red short dashed lines in Fig.~\ref{fig:res_cp_h}.

4) $\theta_0 \rightarrow 0$, $\mathbf{h}_0 = (0, 0, h_{z0})$. This situation characterises the enhancing of the power loss by the static field independently on its value, when $\widetilde{\Omega} \gg h$. In another case, when $\widetilde{\Omega} < h$, the weak enough static field ($h_{z0} < h$) does not affect the power loss (see the blue dashed-dotted and red long dashed lines in Fig.~\ref{fig:res_cp_h}). If $h_{z0} > h$, the static field decreases $\widetilde{Q}$, see the blue dashed-dotted and red dotted lines in Fig.~\ref{fig:res_cp_h}.

Now let consider the results for action of the linearly polarized field. For better comparability of the obtained results, let define the linearly polarized field using the expression (\ref{eq:def_h_cp}), when $\sigma = 0$. The results of the simulations set as 3D waterfall plots are shown in Fig.~\ref{fig:Res_lp}. Firstly, here the dependence on the initial conditions takes place for all parameters of the alternating field, while in the previous case the mode of uniform rotation is not sensitive to them. Also, the linear polarization of the alternating field leads to additional dependence on initial azimuthal angle $\varphi_0$. But, actually, this fact does not carry any new effects, and further we will always assume that $\varphi_0 = 0$. When the external field oscillates along the direction, which is perpendicular to the initial position of $\mathbf{m}$, or $\theta_0 \rightarrow 0$, the power loss grows monotonously up to the values, predicted by (\ref{eq:Q_hf}). In the opposite case, $\theta_0 \rightarrow \pi/2$, when the alternating field is applied along the magnetic moment in the initial position, the dependencies $\widetilde{Q}(\widetilde{\Omega})$ are unimodal and the condition $\widetilde{Q} \rightarrow 0 $ holds for large enough $\widetilde{\Omega}$. This can be explained by small value of the torque, which  acts on the magnetic moment in the initial state when field of the type $\mathbf{h}(\widetilde{t}) = \mathbf{e}_x h \cos(\widetilde{\Omega} \widetilde{t})$ is applied. Thus, the nanoparticle keeps the initial position and does not responses on the alternating field.

\begin{figure}
    \centering
    \includegraphics {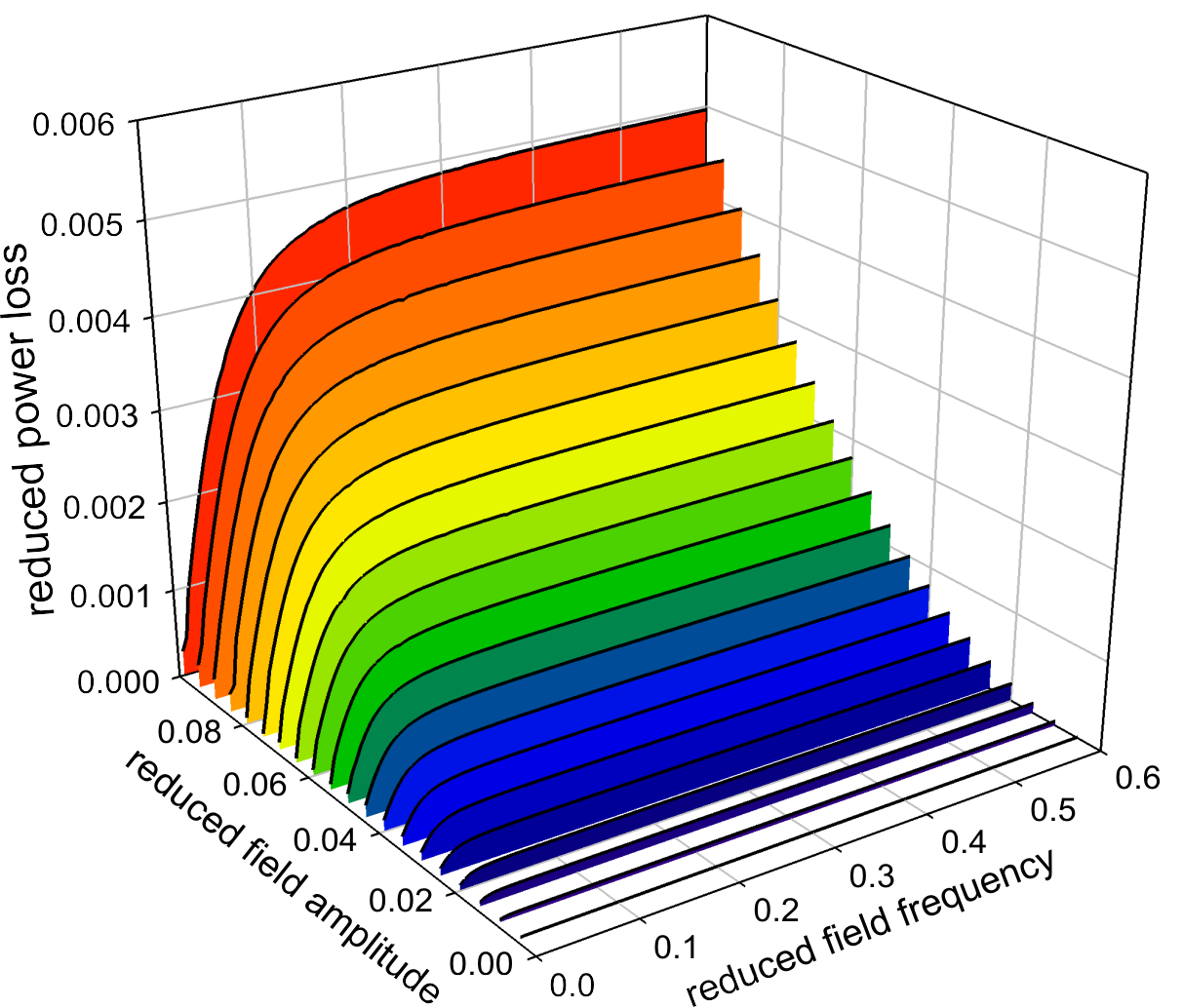}
    \includegraphics {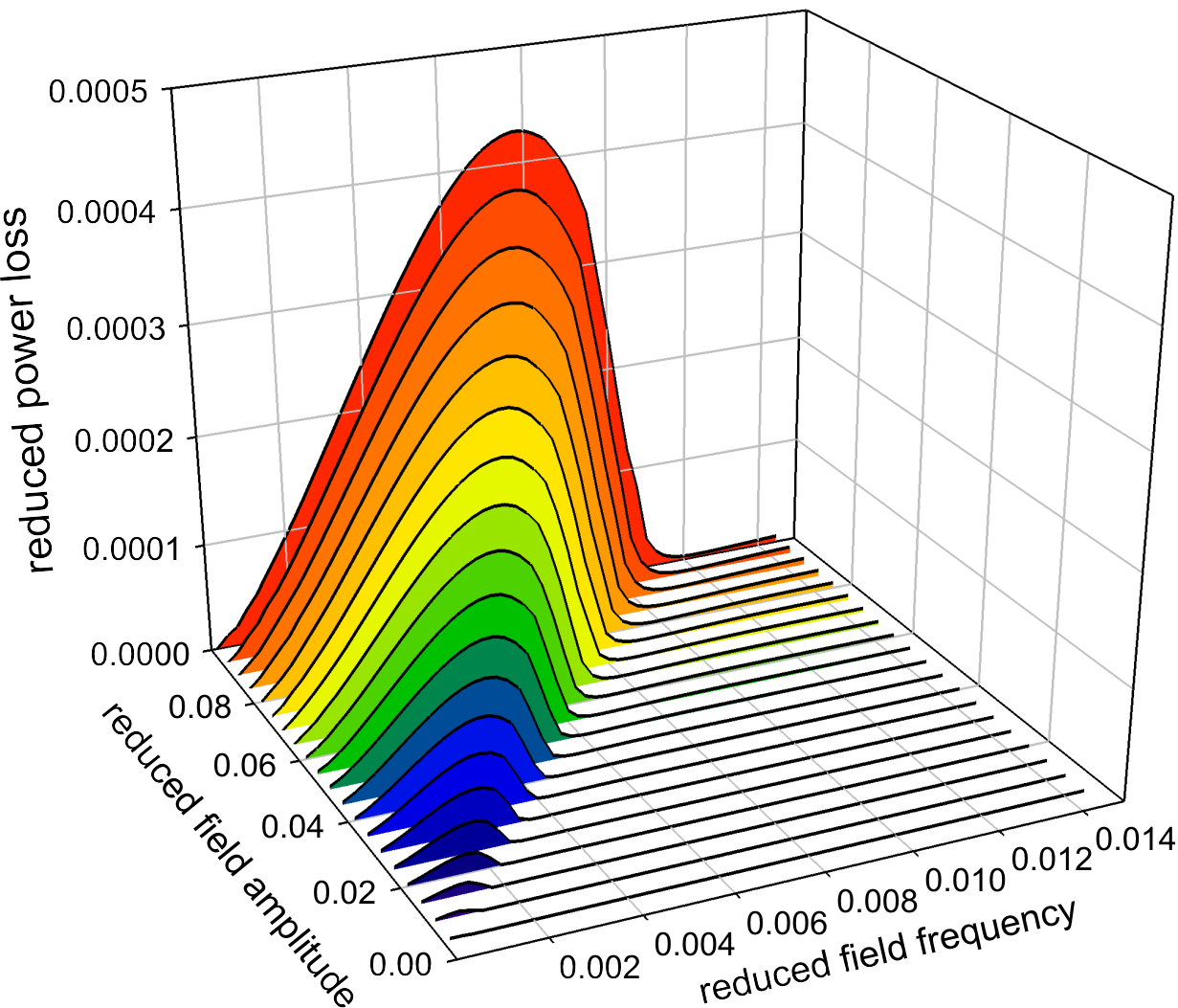}
    \caption {\label{fig:res_lp} (Color online)
    The simulation results for the power loss obtained for different initial conditions of the magnetic moment: $\theta_0 = 0.01$ (a), $\theta_0 = \pi/2-0.01$ (b). Linearly polarized field (\ref{eq:def_h_cp}) ($\sigma = 0$) is acting. The values of the system parameters are chosen as $R=20\mathrm{nm}$, $\mathcal{M} = 3.38\cdot10^{5} \mathrm{A/m}$, $\eta = 5\cdot10^{-3} \mathrm{Pa}\cdot\mathrm{s}^{1}$, $h = 0.05$. The strong dependence on the initial conditions of the magnetic moment is observed.}
\end{figure}

\begin{figure}
    \centering
    \includegraphics {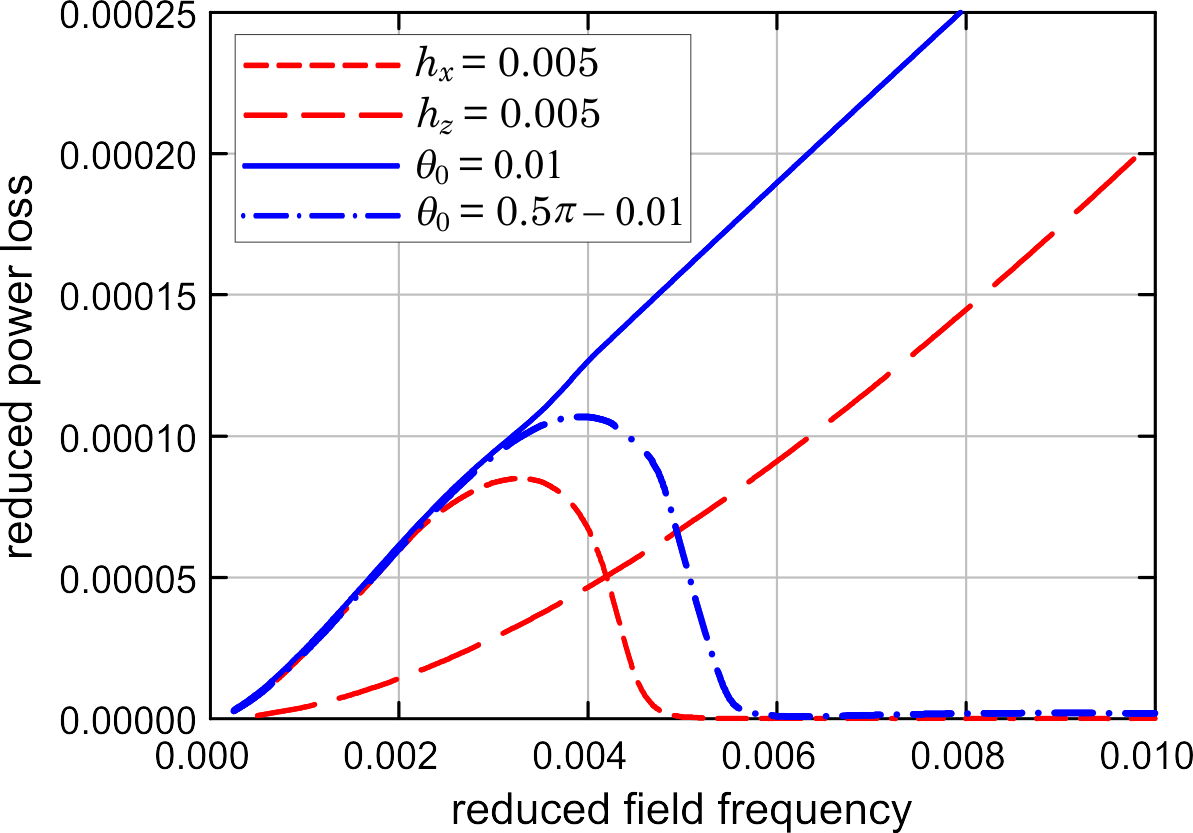}
    \caption {\label{fig:res_lp_h} (Color online)
    The comparison of the numerical results for the power loss obtained for different initial conditions and additional static fields. Linearly polarized field (\ref{eq:def_h_cp}) ($\sigma = 0$) is acting. The values of the system parameters are chosen as $R=20\mathrm{nm}$, $\mathcal{M} = 3.38\cdot10^{5} \mathrm{A/m}$, $\eta = 5\cdot10^{-3} \mathrm{Pa}\cdot\mathrm{s}^{1}$, $h = 0.05$. The possibility of control of the heating rate by a static field is suggested.}
\end{figure}

The influence of the static field on the power loss here also depends strongly on the direction of this field with respect to the initial position of the magnetic moment and the linearly polarized field. The cases of coincidence of the initial positions and static fields ($\theta_0 \rightarrow 0$, $\mathbf{h}_0 = (0,0,h_{z0})$ and $\theta_0 \rightarrow \pi/2$, $\mathbf{h}_0 = (h_{x0},0,0)$) are described in Fig~\ref{fig:res_lp_h}. One can see that the static field reduces the power loss here, because it suppress of the response of $\mathbf{m}$ on the field $\mathbf{h}(\widetilde{t}) = \mathbf{e}_x h \cos(\widetilde{\Omega} \widetilde{t})$ (see the positions of the blue dashed-dotted with respect to the red short-dashed lines and the positions of the blue solid with respect to the red long-dashed lines in Fig~\ref{fig:res_lp_h}). The increase in the $\widetilde{Q}$ value due to the static field is possible for not very small frequencies, when the initial position of $\mathbf{m}$ is close to the line of $\mathbf{h}$ action ($\theta_0 \rightarrow \pi/2$), but the static field tries to increase the angle between them (see the red long-dashed and blue dashed-dotted lines in Fig~\ref{fig:res_lp_h}).

It should be concluded that the static field can increase the power loss when it is applied perpendicularly to the alternating field. This fact gives us another possibility to control the heating rate.

\begin{figure}
    \centering
    \includegraphics {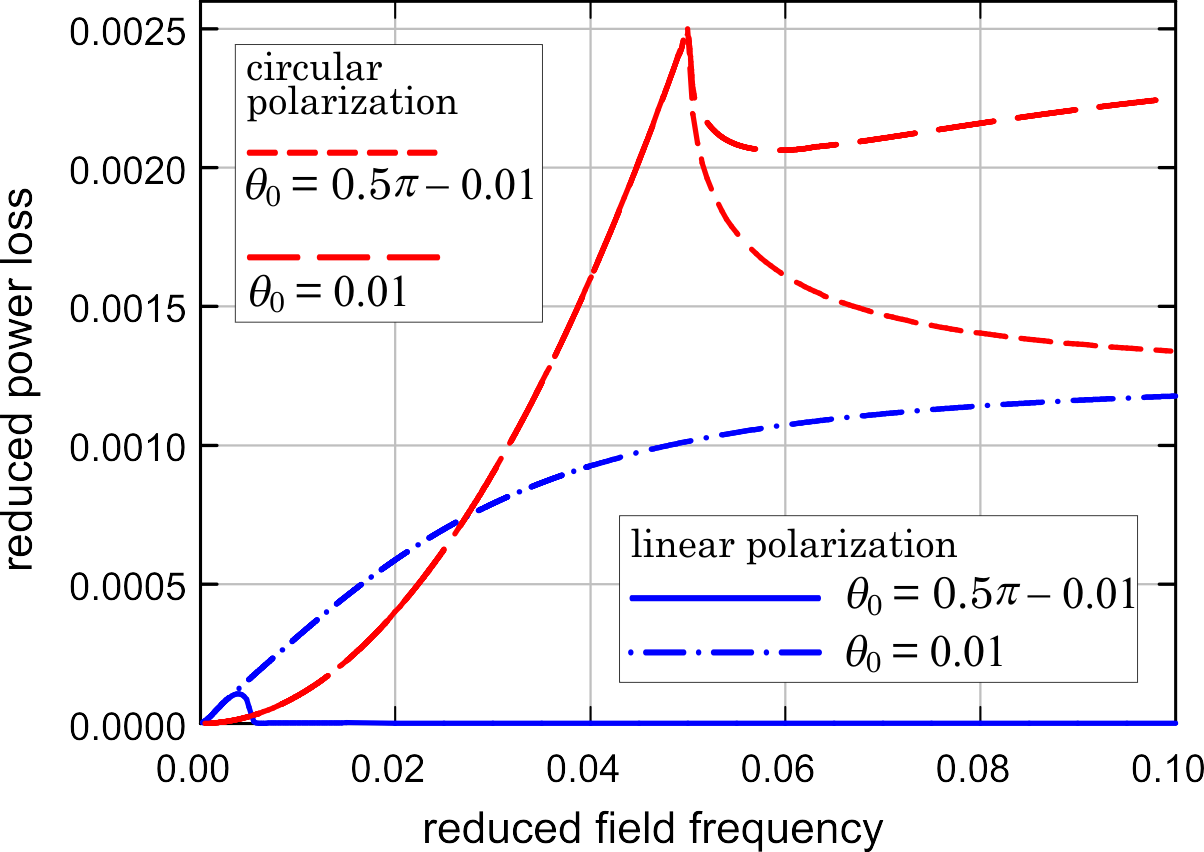}
    \caption {\label{fig:res_cp_lp} (Color online)
    The comparison of the numerical results for the power loss obtained for different polarizations of an alternating field. The values of the system parameters are chosen as $R=20\mathrm{nm}$, $\mathcal{M} = 3.38\cdot10^{5} \mathrm{A/m}$, $\eta = 5\cdot10^{-3} \mathrm{Pa}\cdot\mathrm{s}^{1}$, $h = 0.05$. The advantage of the linearly polarized field for small frequencies is confirmed.}
\end{figure}

The comparison of the power losses for different field polarization is shown in Fig.~\ref{fig:res_cp_lp}. It is seen, that for small frequencies the linear polarization yields considerably larger power loss for the same field parameters and initial conditions. This rather unexpected result can be explained in the following way. While the circularly polarized field rotates the nanoparticle magnetic moment with the field frequency, the linearly polarized field inverts $\mathbf{m}$ direction twice during the field period. The velocity of latter reorientation motion, in fact, is defined by the characteristic time of the nanoparticle response on the external field. In the given case it is smaller than the field period. Thus, the reorientation occurs faster than the field is changed that leads to larger energy consumption correspondingly. When the frequency grows the field period becomes smaller than the characteristic time of the nanoparticle response, and $\mathbf{m}$ does not have time for complete reorientation now. Therefore the circularly polarized field produces larger energy losses in comparison with the linearly polarized one. In other cases for the heating purposes the advantage of the circularly polarized field over the linearly polarized one is obvious and follows from the analysis, presented above.

\section{SUMMARY AND CONCLUSIONS}
\label{Summ}

The deterministic spherical motion of the ferromagnetic nanoparticle in a viscous liquid has been investigated using the angular momentum equation and the equations of spherical motion. The power loss resulted from viscous friction of such nanoparticle driven by the alternating magnetic fields has been studied in detail both analytically and numerically. The circular and linear polarizations of the alternating field have been considered and the heating process has been analyzed in the terms of features of the nanoparticle trajectory.

The circularly polarized field of dimensionless amplitude $h$ and dimensionless frequency $\widetilde{\Omega}$ can generate three periodic modes: uniform precession ($\widetilde{\Omega} \leq h$), non-uniform precession ($\widetilde{\Omega} \sim h$) and rotational oscillations mode ($\widetilde{\Omega} \gg h$). The first and the last modes have been described analytically. The behaviour of the dimensionless power loss is the following: $\widetilde{Q} \sim h^2\widetilde{\Omega}^2$ in uniform mode, $\widetilde{Q} \sim h^2 (1 + \cos^2\theta_0)$ in rotational oscillations mode ($\theta_0$ is the initial polar angle). The non-uniform mode generation is accompanied with decreasing $\widetilde{Q}$. It is important to note the dependence of $\widetilde{Q}$ on the initial conditions and a tendency to saturation while $\widetilde{\Omega} > h$.

The nanoparticle driven by the linearly polarized field performs the oscillations with the field frequency in the defined plane. This motion has been described analytically in the approximation of negligibly small momentum of inertia. The dependence of the power loss on the frequency in a general case is unimodal and tends to constant $\sim h^2 \cos^2\theta_0$. In particular, $\widetilde{Q} \rightarrow 0$ when $\mathbf{m}(0) \parallel \mathbf{h}$, and the energy of linearly polarized field in this case is not absorbed by the nanoparticle ($\mathbf{m}(0)$ is the unit vector of magnetic moment in the initial state).

The essential impact of the static field $\mathbf{h}_0$ on the process of transformation of the alternating field energy into the heat one has been analyzed. If the static field strive to direct the magnetic moment perpendicularly to the plane (line) of the alternating field action, it increases the power loss. At the same time, $\mathbf{h}_0$ bonds the magnetic moment and decreases its response on the alternating filed, that leads to the reduction of the power loss. As a result of the competition of these two effects, the static filed can enhance the heating efficiency in the following situations. 1) The circularly polarized field is applied,  $\widetilde{\Omega} > h$ ($\widetilde{\Omega} \sim h$), $\mathbf{h}_0$ to the polarization plane, $h_0 \ll h$. 2) The linearly polarized field is applied, $\widetilde{\Omega} > h_0$, $\mathbf{h}_0 \perp \mathbf{h}$, $\theta_0 \approx \pi/2$. 3) If $\widetilde{\Omega} \gg h$, $\theta_0 = \pi/2$, the values of $\widetilde{Q}$ can be twice as large, for the any polarization type of external alternating field $\mathbf{h}$ and any values of external static field $\mathbf{h}_0$ while $\mathbf{h} \perp \mathbf{h}_0$.

Finally, it should be concluded that linearly polarized field is more efficient for heating in comparison with the circularly polarized one, while frequency is small enough ($\widetilde{\Omega} \ll h$). The origin of such behavior is in the fast reorientations of the magnetic moments between two antiparallel direction, caused by the linearly polarized field. In the rest range of frequencies, the rotational field characterized by the power loss up to two times larger ($\widetilde{\Omega} \gg h$) for the same initial conditions.

\section*{ACKNOWLEDGMENTS}

The author express appreciation to S.~I.~Denisov and Yu.~S.~Bystrik for the valuable comments and discussion. The author is also grateful to the Ministry of Education and Science of Ukraine for partial financial support under Grant No.~0116U002622.


\bibliography{Lyutyy_RigidDipole_PowerLoss}

\begin{thebibliography}{10}
\expandafter\ifx\csname url\endcsname\relax
  \def\url#1{\texttt{#1}}\fi
\expandafter\ifx\csname urlprefix\endcsname\relax\def\urlprefix{URL }\fi
\expandafter\ifx\csname href\endcsname\relax
  \def\href#1#2{#2} \def\path#1{#1}\fi

\bibitem{Rosensweig1985Ferrohydrodynamics}
R.~Rosensweig, Ferrohydrodynamics, Cambridge University Press, 1985.

\bibitem{0038-5670-17-2-R02}
M.~I. Shliomis, \href{http://stacks.iop.org/0038-5670/17/i=2/a=R02}{Magnetic
  fluids}, Soviet Physics Uspekhi 17~(2) (1974) 153.
\newline\urlprefix\url{http://stacks.iop.org/0038-5670/17/i=2/a=R02}

\bibitem{0022-3727-36-13-201}
Q.~A. Pankhurst, J.~Connolly, S.~K. Jones, J.~Dobson,
  \href{http://stacks.iop.org/0022-3727/36/i=13/a=201}{Applications of magnetic
  nanoparticles in biomedicine}, Journal of Physics D: Applied Physics 36~(13)
  (2003) R167.
\newline\urlprefix\url{http://stacks.iop.org/0022-3727/36/i=13/a=201}

\bibitem{VEISEH2010284}
O.~Veiseh, J.~W. Gunn, M.~Zhang,
  \href{http://www.sciencedirect.com/science/article/pii/S0169409X09003408}{Design
  and fabrication of magnetic nanoparticles for targeted drug delivery and
  imaging}, Advanced Drug Delivery Reviews 62~(3) (2010) 284 -- 304, targeted
  Delivery Using Inorganic Nanosystem.
\newblock \href {http://dx.doi.org/https://doi.org/10.1016/j.addr.2009.11.002}
  {\path{doi:https://doi.org/10.1016/j.addr.2009.11.002}}.
\newline\urlprefix\url{http://www.sciencedirect.com/science/article/pii/S0169409X09003408}

\bibitem{JORDAN1999413}
A.~Jordan, R.~Scholz, P.~Wust, H.~Fähling, R.~Felix,
  \href{http://www.sciencedirect.com/science/article/pii/S0304885399000888}{Magnetic
  fluid hyperthermia (mfh): Cancer treatment with ac magnetic field induced
  excitation of biocompatible superparamagnetic nanoparticles}, Journal of
  Magnetism and Magnetic Materials 201~(1) (1999) 413 -- 419.
\newblock \href
  {http://dx.doi.org/http://dx.doi.org/10.1016/S0304-8853(99)00088-8}
  {\path{doi:http://dx.doi.org/10.1016/S0304-8853(99)00088-8}}.
\newline\urlprefix\url{http://www.sciencedirect.com/science/article/pii/S0304885399000888}

\bibitem{0957-4484-25-45-452001}
S.~Dutz, R.~Hergt,
  \href{http://stacks.iop.org/0957-4484/25/i=45/a=452001}{Magnetic particle
  hyperthermia: promising tumour therapy?}, Nanotechnology 25~(45) (2014)
  452001.
\newline\urlprefix\url{http://stacks.iop.org/0957-4484/25/i=45/a=452001}

\bibitem{TIAN2016420}
B.~Tian, Z.~Qiu, J.~Ma, T.~Z.~G. de~la Torre, C.~Johansson, P.~Svedlindh,
  M.~Strömberg,
  \href{http://www.sciencedirect.com/science/article/pii/S095656631630625X}{Attomolar
  zika virus oligonucleotide detection based on loop-mediated isothermal
  amplification and ac susceptometry}, Biosensors and Bioelectronics 86 (2016)
  420 -- 425.
\newblock \href {http://dx.doi.org/https://doi.org/10.1016/j.bios.2016.06.085}
  {\path{doi:https://doi.org/10.1016/j.bios.2016.06.085}}.
\newline\urlprefix\url{http://www.sciencedirect.com/science/article/pii/S095656631630625X}

\bibitem{C6AY00721J}
B.~Tian, T.~Zardan Gomez de~la Torre, M.~Donolato, M.~F. Hansen, P.~Svedlindh,
  M.~Stromberg, \href{http://dx.doi.org/10.1039/C6AY00721J}{Multi-scale
  magnetic nanoparticle based optomagnetic bioassay for sensitive dna and
  bacteria detection}, Anal. Methods 8 (2016) 5009--5016.
\newblock \href {http://dx.doi.org/10.1039/C6AY00721J}
  {\path{doi:10.1039/C6AY00721J}}.
\newline\urlprefix\url{http://dx.doi.org/10.1039/C6AY00721J}

\bibitem{0022-3727-39-22-002}
H.~Xi, K.-Z. Gao, Y.~Shi, S.~Xue,
  \href{http://stacks.iop.org/0022-3727/39/i=22/a=002}{Precessional dynamics of
  single-domain magnetic nanoparticles driven by small ac magnetic fields},
  Journal of Physics D: Applied Physics 39~(22) (2006) 4746.
\newline\urlprefix\url{http://stacks.iop.org/0022-3727/39/i=22/a=002}

\bibitem{doi:10.1063/1.4737126}
N.~A. Usov, B.~Y. Liubimov, \href{http://dx.doi.org/10.1063/1.4737126}{Dynamics
  of magnetic nanoparticle in a viscous liquid: Application to magnetic
  nanoparticle hyperthermia}, Journal of Applied Physics 112~(2) (2012) 023901.
\newblock \href {http://arxiv.org/abs/http://dx.doi.org/10.1063/1.4737126}
  {\path{arXiv:http://dx.doi.org/10.1063/1.4737126}}, \href
  {http://dx.doi.org/10.1063/1.4737126} {\path{doi:10.1063/1.4737126}}.
\newline\urlprefix\url{http://dx.doi.org/10.1063/1.4737126}

\bibitem{PhysRevB.95.134447}
H.~Keshtgar, S.~Streib, A.~Kamra, Y.~M. Blanter, G.~E.~W. Bauer,
  \href{https://link.aps.org/doi/10.1103/PhysRevB.95.134447}{Magnetomechanical
  coupling and ferromagnetic resonance in magnetic nanoparticles}, Phys. Rev. B
  95 (2017) 134447.
\newblock \href {http://dx.doi.org/10.1103/PhysRevB.95.134447}
  {\path{doi:10.1103/PhysRevB.95.134447}}.
\newline\urlprefix\url{https://link.aps.org/doi/10.1103/PhysRevB.95.134447}

\bibitem{doi:10.21272/jnep.8(4(2)).04086}
T.~V. Lyutyy, O.~M. Hryshko, A.~A. Kovner, E.~S. Denisova,
  \href{https://dx.doi.org/10.21272/jnep.8(4(2)).04086}{Precession of a fine
  magnetic particle with finite anisotropy in a viscous fluid}, J. Nano-
  Electron. Phys. 8~(4) (2016) 04086.
\newblock \href {http://dx.doi.org/10.21272/jnep.8(4(2)).04086}
  {\path{doi:10.21272/jnep.8(4(2)).04086}}.
\newline\urlprefix\url{https://dx.doi.org/10.21272/jnep.8(4(2)).04086}

\bibitem{Lyutyy201887}
T.~V. Lyutyy, O.~M. Hryshko, A.~A. Kovner,
  \href{http://www.sciencedirect.com/science/article/pii/S0304885317310740}{Power
  loss for a periodically driven ferromagnetic nanoparticle in a viscous fluid:
  The finite anisotropy aspects}, Journal of Magnetism and Magnetic Materials
  446~(Supplement C) (2018) 87 -- 94.
\newblock \href {http://dx.doi.org/https://doi.org/10.1016/j.jmmm.2017.09.021}
  {\path{doi:https://doi.org/10.1016/j.jmmm.2017.09.021}}.
\newline\urlprefix\url{http://www.sciencedirect.com/science/article/pii/S0304885317310740}

\bibitem{PhysRevB.95.104430}
K.~D. Usadel,
  \href{https://link.aps.org/doi/10.1103/PhysRevB.95.104430}{Dynamics of
  magnetic nanoparticles in a viscous fluid driven by rotating magnetic
  fields}, Phys. Rev. B 95 (2017) 104430.
\newblock \href {http://dx.doi.org/10.1103/PhysRevB.95.104430}
  {\path{doi:10.1103/PhysRevB.95.104430}}.
\newline\urlprefix\url{https://link.aps.org/doi/10.1103/PhysRevB.95.104430}

\bibitem{0031-9155-63-3-035004}
J.~Weizenecker, \href{http://stacks.iop.org/0031-9155/63/i=3/a=035004}{The
  fokker-planck equation for coupled brown-n\'{e}el-rotation}, Physics in
  Medicine and Biology 63~(3) (2018) 035004.
\newline\urlprefix\url{http://stacks.iop.org/0031-9155/63/i=3/a=035004}

\bibitem{PhysRevE.63.011504}
C.~Scherer, H.-G. Matuttis,
  \href{https://link.aps.org/doi/10.1103/PhysRevE.63.011504}{Rotational
  dynamics of magnetic particles in suspensions}, Phys. Rev. E 63 (2000)
  011504.
\newblock \href {http://dx.doi.org/10.1103/PhysRevE.63.011504}
  {\path{doi:10.1103/PhysRevE.63.011504}}.
\newline\urlprefix\url{https://link.aps.org/doi/10.1103/PhysRevE.63.011504}

\bibitem{PhysRevE.83.021401}
Y.~L. Raikher, V.~I. Stepanov,
  \href{https://link.aps.org/doi/10.1103/PhysRevE.83.021401}{Power losses in a
  suspension of magnetic dipoles under a rotating field}, Phys. Rev. E 83
  (2011) 021401.
\newblock \href {http://dx.doi.org/10.1103/PhysRevE.83.021401}
  {\path{doi:10.1103/PhysRevE.83.021401}}.
\newline\urlprefix\url{https://link.aps.org/doi/10.1103/PhysRevE.83.021401}

\bibitem{0953-8984-15-23-313}
B.~U. Felderhof, R.~B. Jones,
  \href{http://stacks.iop.org/0953-8984/15/i=23/a=313}{Mean field theory of the
  nonlinear response of an interacting dipolar system with rotational diffusion
  to an oscillating field}, Journal of Physics: Condensed Matter 15~(23) (2003)
  4011.
\newline\urlprefix\url{http://stacks.iop.org/0953-8984/15/i=23/a=313}

\bibitem{SotoAquino201546}
D.~Soto-Aquino, C.~Rinaldi,
  \href{http://www.sciencedirect.com/science/article/pii/S0304885315301335}{Nonlinear
  energy dissipation of magnetic nanoparticles in oscillating magnetic fields},
  Journal of Magnetism and Magnetic Materials 393 (2015) 46 -- 55.
\newblock \href {http://dx.doi.org/https://doi.org/10.1016/j.jmmm.2015.05.009}
  {\path{doi:https://doi.org/10.1016/j.jmmm.2015.05.009}}.
\newline\urlprefix\url{http://www.sciencedirect.com/science/article/pii/S0304885315301335}

\bibitem{PhysRevE.92.042312}
T.~V. Lyutyy, S.~I. Denisov, V.~V. Reva, Y.~S. Bystrik,
  \href{https://link.aps.org/doi/10.1103/PhysRevE.92.042312}{Rotational
  properties of ferromagnetic nanoparticles driven by a precessing magnetic
  field in a viscous fluid}, Phys. Rev. E 92 (2015) 042312.
\newblock \href {http://dx.doi.org/10.1103/PhysRevE.92.042312}
  {\path{doi:10.1103/PhysRevE.92.042312}}.
\newline\urlprefix\url{https://link.aps.org/doi/10.1103/PhysRevE.92.042312}

\bibitem{PhysRevE.97.052611}
T.~V. Lyutyy, V.~V. Reva,
  \href{https://link.aps.org/doi/10.1103/PhysRevE.97.052611}{Energy dissipation
  of rigid dipoles in a viscous fluid under the action of a time-periodic
  field: The influence of thermal bath and dipole interaction}, Phys. Rev. E 97
  (2018) 052611.
\newblock \href {http://dx.doi.org/10.1103/PhysRevE.97.052611}
  {\path{doi:10.1103/PhysRevE.97.052611}}.
\newline\urlprefix\url{https://link.aps.org/doi/10.1103/PhysRevE.97.052611}

\bibitem{doi:10.1063/1.1656014}
J.~J. Newman, R.~B. Yarbrough, \href{https://doi.org/10.1063/1.1656014}{Motions
  of a magnetic particle in a viscous medium}, Journal of Applied Physics
  39~(12) (1968) 5566--5569.
\newblock \href {http://arxiv.org/abs/https://doi.org/10.1063/1.1656014}
  {\path{arXiv:https://doi.org/10.1063/1.1656014}}, \href
  {http://dx.doi.org/10.1063/1.1656014} {\path{doi:10.1063/1.1656014}}.
\newline\urlprefix\url{https://doi.org/10.1063/1.1656014}

\bibitem{andr2006magnetism}
W.~Andr\"{a}, H.~Nowak,
  \href{https://onlinelibrary.wiley.com/doi/book/10.1002/9783527610174}{Magnetism
  in Medicine: A Handbook}, Wiley-VCH Verlag GmbH \& Co. KGaA, 2007.
\newblock \href {http://dx.doi.org/10.1002/9783527610174}
  {\path{doi:10.1002/9783527610174}}.
\newline\urlprefix\url{https://onlinelibrary.wiley.com/doi/book/10.1002/9783527610174}

\bibitem{Frenkel:106808}
Y.~I. Frenkel, \href{https://cds.cern.ch/record/106808}{{Kinetic theory of
  liquids}}, Dover, New York, NY, 1955.
\newline\urlprefix\url{https://cds.cern.ch/record/106808}

\bibitem{PhysRevB.91.054425}
T.~V. Lyutyy, S.~I. Denisov, A.~Y. Peletskyi, C.~Binns,
  \href{http://link.aps.org/doi/10.1103/PhysRevB.91.054425}{Energy dissipation
  in single-domain ferromagnetic nanoparticles: Dynamical approach}, Phys. Rev.
  B 91 (2015) 054425.
\newblock \href {http://dx.doi.org/10.1103/PhysRevB.91.054425}
  {\path{doi:10.1103/PhysRevB.91.054425}}.
\newline\urlprefix\url{http://link.aps.org/doi/10.1103/PhysRevB.91.054425}

\bibitem{C3RA45457F}
M.~I. Dar, S.~A. Shivashankar,
  \href{http://dx.doi.org/10.1039/C3RA45457F}{Single crystalline magnetite{,}
  maghemite{,} and hematite nanoparticles with rich coercivity}, RSC Adv. 4
  (2014) 4105--4113.
\newblock \href {http://dx.doi.org/10.1039/C3RA45457F}
  {\path{doi:10.1039/C3RA45457F}}.
\newline\urlprefix\url{http://dx.doi.org/10.1039/C3RA45457F}

\bibitem{PhysRev.130.1677}
W.~F. Brown, \href{https://link.aps.org/doi/10.1103/PhysRev.130.1677}{Thermal
  fluctuations of a single-domain particle}, Phys. Rev. 130 (1963) 1677--1686.
\newblock \href {http://dx.doi.org/10.1103/PhysRev.130.1677}
  {\path{doi:10.1103/PhysRev.130.1677}}.
\newline\urlprefix\url{https://link.aps.org/doi/10.1103/PhysRev.130.1677}

\bibitem{DENISOV1998282}
S.~Denisov, A.~Yunda,
  \href{http://www.sciencedirect.com/science/article/pii/S0921452697008788}{Thermal-induced
  inversion of the magnetic moment in superparamagnetic particles}, Physica B:
  Condensed Matter 245~(3) (1998) 282 -- 287.
\newblock \href
  {http://dx.doi.org/https://doi.org/10.1016/S0921-4526(97)00878-8}
  {\path{doi:https://doi.org/10.1016/S0921-4526(97)00878-8}}.
\newline\urlprefix\url{http://www.sciencedirect.com/science/article/pii/S0921452697008788}

\bibitem{PhysRevLett.97.227202}
S.~I. Denisov, T.~V. Lyutyy, P.~H\"anggi,
  \href{https://link.aps.org/doi/10.1103/PhysRevLett.97.227202}{Magnetization
  of nanoparticle systems in a rotating magnetic field}, Phys. Rev. Lett. 97
  (2006) 227202.
\newblock \href {http://dx.doi.org/10.1103/PhysRevLett.97.227202}
  {\path{doi:10.1103/PhysRevLett.97.227202}}.
\newline\urlprefix\url{https://link.aps.org/doi/10.1103/PhysRevLett.97.227202}

\bibitem{Goldstein}
H.~Goldstein,
  \href{https://books.google.com.ua/books?id=Spy6xHWFJIEC}{Classical
  Mechanics}, Pearson Education, 2002.
\newline\urlprefix\url{https://books.google.com.ua/books?id=Spy6xHWFJIEC}

\bibitem{doi:10.1142/7861}
M.~Gitterman, \href{https://www.worldscientific.com/doi/abs/10.1142/7861}{The
  Chaotic Pendulum}, World Scientific, 5 Toh Tuck Link, Singapore 596224, 2010.
\newblock \href
  {http://arxiv.org/abs/https://www.worldscientific.com/doi/pdf/10.1142/7861}
  {\path{arXiv:https://www.worldscientific.com/doi/pdf/10.1142/7861}}, \href
  {http://dx.doi.org/10.1142/7861} {\path{doi:10.1142/7861}}.
\newline\urlprefix\url{https://www.worldscientific.com/doi/abs/10.1142/7861}

\bibitem{Raikher_1994}
Y.~L. Raikher, M.~I. Shliomis,
  \href{http://dx.doi.org/10.1002/9780470141465.ch8}{The effective field method
  in the orientational kinetics of magnetic fluids and liquid crystals},
  Advances in Chemical Physics 87 (1994) 595--751.
\newblock \href {http://dx.doi.org/http://dx.doi.org/10.1002/9780470141465.ch8}
  {\path{doi:http://dx.doi.org/10.1002/9780470141465.ch8}}.
\newline\urlprefix\url{http://dx.doi.org/10.1002/9780470141465.ch8}

\bibitem{PhysRevLett.86.724}
G.~Bertotti, C.~Serpico, I.~D. Mayergoyz,
  \href{https://link.aps.org/doi/10.1103/PhysRevLett.86.724}{Nonlinear
  magnetization dynamics under circularly polarized field}, Phys. Rev. Lett. 86
  (2001) 724--727.
\newblock \href {http://dx.doi.org/10.1103/PhysRevLett.86.724}
  {\path{doi:10.1103/PhysRevLett.86.724}}.
\newline\urlprefix\url{https://link.aps.org/doi/10.1103/PhysRevLett.86.724}

\bibitem{0953-8984-21-39-396002}
T.~V. Lyutyy, A.~Y. Polyakov, A.~V. Rot-Serov, C.~Binns,
  \href{http://stacks.iop.org/0953-8984/21/i=39/a=396002}{Switching properties
  of ferromagnetic nanoparticles driven by a circularly polarized magnetic
  field}, Journal of Physics: Condensed Matter 21~(39) (2009) 396002.
\newline\urlprefix\url{http://stacks.iop.org/0953-8984/21/i=39/a=396002}

\end{thebibliography}

\end{document}